\documentclass[aps,prc,floatfix,12pt]{revtex4}

\usepackage{epsfig}
\usepackage{amsmath}

\begin{document}

\title{Initiatives in Nuclear Theory at the Variable Energy Cyclotron Centre}
\author{D. K. Srivastava{\footnote{Lecture given in the special
session, ``Reflections'' during the Workshop on Changing Scales in
Nuclear Physics at Variable Energy Cyclotron Centre, Kolkata,
 June 14-15, 2005, to honour Prof. Bikash Sinha on his 60$^{\mathrm{th}}$ birth-day. }},
J. Alam, D. N. Basu, A. K. Chaudhuri, J. N. De,  K. Krishan, S. Pal}

\affiliation{Variable Energy Cyclotron 
Centre, 1/AF Bidhan Nagar, Kolkata 700 064, India}            

\date{\today}

\begin{abstract}

We recall the path breaking contributions of the nuclear theory group of 
the Variable Energy Cyclotron Centre, Kolkata. From a beginning of just one person
 in 1970s, the group has steadily developed into a leading group in
 the country today, with seminal contributions to almost the entire
 range of nuclear physics, viz., low energy nuclear reactions, 
nuclear structure, deep inelastic collisions, fission, liquid to
 gas phase transitions, nuclear matter, equation of state, mass formulae,
 neutron stars, relativistic heavy ion collisions, medium modification of 
hadron properties, quark gluon plasma, and cosmology of early universe.

\end{abstract}

\maketitle
\section{Prologue}

Research in nuclear theory remains one of the most challenging and rewarding
intellectual pursuits today. Years of hard and inspired work by thousands 
of scientists from across the world over the last century has helped us
 understand the structure of nucleons, nuclei, neutron stars, nuclear matter, 
and nuclear forces. A stage has reached when experimentally determined
 nucleon-nucleon interaction can be directly used to calculate the structure
 of nuclei from first principles. Speculations about the existence of
a de-confined strongly interacting matter under conditions of extreme 
temperatures are on the verge of being fully confirmed. Modifications
 of hadronic properties under extreme conditions, not encountered since
 the Big Bang or outside the core of neutron stars will be fully 
investigated in coming years. Nuclei- very far from the line of 
beta stability  will be studied, which will definitely have very 
unusual properties. Remnants of the Big Bang will be searched for and 
constituents of dark matter will be identified.  All these efforts 
will be firmly rooted in reliable, accurate, and precise nuclear theory,
where almost all calculations will be done from first principles. 
These results are a remarkable embodiment of the vision that elements
 of reality can be reproduced from purely conceptual constructions.
 This heralds the ultimate success: to be able to understand the world 
fully in terms of only intellect. Today we stand at a historical moment.
 It is always interesting and rewarding,
to look back and reflect upon the path, which 
transported us to this moment in intellectual space and time.
It also helps recognize the evolution of a thought process and a 
subject `engineered' by us over years.
 This is 
also to put on record the valiant efforts of the persons who initiated
 the activities against every imaginable and impossible odds and 
deprivation and who preserved and persisted under the face of technology
denial and too small a number necessary to sustain a truly critical system.

It is a matter of considerable satisfaction that the nuclear theory 
group at the cyclotron centre has made several path-breaking contributions 
and has consistently provided valuable help in planning and execution of
 our nuclear physics programme. This, rather subjective, narrative is 
an attempt to recall various interesting contributions of the group, which
 started functioning in 1970, with just one person, and an availability 
of a computer time of just a few hours per month, on a computer 
called CDC-3600, which had a total memory of 32 K-words and a speed which
 was perhaps one million times less than that of a personal computer 
available today for computer games to be played by children.
 
\section{The Genesis}

Dr. Nripendra Kumar Ganguly joined the Variable Energy Cyclotron "Project" 
at the Bhabha Atomic Research Centre, at Trombay in mid 1970 as a
 Project Officer. He was a student of legendary Profs. D. M. Bose, 
S. N. Bose, M. N. Saha, and B. D. Nagchaudhury of University of Calcutta.
 He had obtained  a Ph. D. from North Carolina State University, USA,
where he obtained the most accurate value for the diffusion
coefficient of heavy water, needed for the design of nuclear
reactors.   He worked at the Carnegie Melon Institute (now 
University) in USA, at the Institute of Nuclear Physics, Orsay,
and at the Rutherford High Energy Laboratory, in England, before 
returning to India. He worked briefly at the
 Aligrah Muslim University before joining VECC.

 It was his desire to initiate state of the art
 calculations of the nuclear theory and to have
all the computer codes for the calculations of single particle
 bound state calculations\cite{first},
 shell-model calculations, direct nuclear reactions, 
coupled reaction channels~\cite{codes}, etc., several of which he had 
developed while in England.

Experiments for these studies were to be performed at the Variable
 Energy Cyclotron being built at Calcutta. He was of the opinion 
that our ability to provide a precise explanation and prediction 
of experimental data was the only proof that we have understood 
the nuclear structure and the behaviour of nuclear forces. He "borrowed"
Shashi Ranjan Pandey, a student of Prof. K. Ramareddy, from Aligarh 
Muslim University and started work on excited states of $^{4}$He, $^{5}$He
 and $^{6}$He. It was more than 20 years before
the studies of radioactive nuclei were to become fashionable! 

Dr Ganguly  wanted to plan and perform experiments where, for example even
 a simple elastic scattering study of say
protons or alpha particles could be taken to a great height of 
sophistication and artistry.  He talked of the possibilities of 
an on-line analysis of data measured in the experiment.  He suggested
 that one could then use an optical model calculation to predict 
regions of phase space where data changed rapidly and were most 
sensitive to the parameters used, and make a "mid-course correction" 
of the planned experiment to perform measurements there, in an
 iterative manner! The concept of "simulation" was not even
 discussed in the literature at that time, when on the basis of 
calculations he planned to take measurements in very fine angular 
or energy steps in some regions and only in large steps in other
regions! 

He tried to initiate research in algebraic programming,
 when tools like Mathematica were still decades away. He talked of 
tracking cascades generated by a high-energy proton traveling 
through a nucleus at a high energy and studying its thermalization. 
Event generators of high-energy physics were still beyond the horizon.
Shashi Ranjan Pandey was rather ``scared'' of these ``visions'' and 
``fled'' to USA within two years!

 Dr. Ganguly was truly responsible for ushering in an era of large
 scale computing in nuclear physics in the country, by not only 
campaigning for but also taking up the responsibility of the 
installation and running of BESM-6 and IRIS-80 computers in Bombay 
and Calcutta respectively. 

It is worth while to recall his other
visions, which did materialize in course of time, viz., installation of
an ECR heavy ion source at the VECC (now operating at VECC)
, construction of a separated
sector cyclotron  to be coupled to the VECC (it is now being discussed
in connection with the accelerator driven sub-critical systems), a
super-conducting cyclotron (now under construction at Kolkata), and 
the installation of a pelletron in the university system (
which was to become Nuclear Science Centre, New Delhi).

 Dinesh Kumar Srivastava joined him in August 1971. Dr. Ganguly 
gave a course of lectures on nuclear structure, scattering theory,
 nuclear reactions, distorted waves Born approximation, distorted 
waves impulse approximation, etc. to a group of just three persons- 
the second student Ms. Sheela Roy had joined by then. These lectures were
invariably in the evening in a small room in Modular Laboratories, after 
most of the people had left.

In order to kick-start the research activity in the fledgeling group, 
Dr. Ganguly invited Dr.  Bikash Sinha, who was working at King's College,
 London for a period of three months, in July 1972. This visit was
 to have a most profound effect on the group and the future of 
basic research in the Department of Atomic Energy. Dr. Sinha had 
just then published several pioneering papers on the folding model 
for optical model potential, where for the
first time experimentally measured proton densities and 
numerically calculated neutron densities were folded with a 
density dependent nucleon- nucleon interaction to get the
optical model potential which provided precise explanation of 
elastic scattering data for protons.  This method remains un-altered today,
 more than thirty years after it was enunciated and is being used to
 study scattering of radioactive nuclei.  The first journal
publication of the VECC was a result of this visit~\cite{first_j}.

 Sheela Roy (she became Mukhopadhyay in course of time),
 D. K. Srivastava, and several years later D. N. Basu, 
and A. K.  Chaudhuri were to obtain their Ph. D. working on optical
 model. Santanu Pal, who joined in August 1973, also worked on Optical
 Model in the beginning- three of his important observations were the
 effect of coupling of deuteron stripping reactions to elastic scattering
 of deuterons, an explanation of anomalous large angle scattering of alpha 
particles from $^{40}$Ca, and most importantly, stripping of deuterons 
to unbound (resonant) states of nuclei. The last named study was to 
become the harbinger of break-up of light ions, a topic, which has
 continued to grow in importance, sophistication, and reach.

Dr. Bikash Sinha kept a close contact with the group and continued
 to provide valuable help in every possible manner. He had also moved 
to Bhabha Atomic Research Centre, Bombay by 1976. Santanu Pal took up a
 lead from him and used linear response theory to understand dissipation 
in deep inelastic collisions; a subject, which had become quite
exciting at that time and developed it to a high level of sophistication 
and accuracy- going on to distinguish between one and two body friction 
in nuclear collisions. He has now moved along and tackles the
 dynamics of nuclear fission.

Dr. N. K. Ganguly retired in October 1983 and Dr. Bikash Sinha
 moved to Kolkata in November 1983. This was a turning point in the 
theory activities of VECC. Dr. Jadu Nath De had moved to VECC, a few 
years earlier, first as a visiting scientist and then on a
permanent position. Dr. Kewal Krishan, who had joined VECC a few 
years earlier, as a computer software person, D. N. Basu, and A. K.
 Chaudhury had started working in nuclear theory.
Sailajananda Bhattacharya joined forces with Dr.  J. N. De and
Dr. Kewal Krishan, and performed a large number of intricate and valuable
studies in nucleus-nucleus collisions. We lost him to the
experimentalists though, within a few years!  

By then 
the next generation, first Jan-e Alam, then Sourav
Sarkar, and finally Ms. Gargi Chaudhury joined the theory
group.  The group started a journey on the
new vistas provided by the emerging field of quark gluon plasma,
and its work  by this time 
had started being internationally recognized. These efforts received
a great deal of help from a long series of lectures given by Prof. Binayak
Dutta- Roy.

Even the third generation of the group- represented by 
Jajati Kesari Nayak has started making its presence felt! We have 
also benefited from the transfers of Dr. A. K. Dhara and Dr. T. K.
Mukhopadhyay from Bombay. All these people have launched a vigorous activity
 on the vast canvass covering nuclear theory. A number of Ph. D.
 students and post-doctoral fellows,
as well as collaborators, both from India and abroad, have made
 significant contributions.
 There is a rumour that it is considered a privilege to be a member
of the theory group of the VECC. There is also a rumour that the group
``steals'' the best students!

\section{Initial Successes}

\subsection{Optical Model Studies}
Let us now briefly recall some of the initial successes of the
 fledgeling theory group. It is
important as it finds mention in textbooks of Nuclear Physics
and every review article written around that time.
 The success of the folding
model with the exchange effects, encouraged us to 
take an exhaustive study of
all the available data on proton elastic scattering. 
Thus the optical potential was
calculated for all the nuclei for which data were available. 
Around this time W. D. Myers
suggested that for leptodermous (``having a thin skin'')
 distributions like nuclear densities
and nuclear potentials, the best representative of the radius was
 `` the equivalent sharp radius'',
defined by:
\begin{equation}
\frac{4\pi}{3}R^3 f({\mathrm{bulk}})=4\pi \int r^2 \, dr \, f(r),
\end{equation}
where $f({\mathrm{bulk}})$ is the value for the bulk of the distribution.
 With this definition, it was
found that the equivalent sharp radii of the nuclear optical model
 potentials for protons
($R_V$) and the nuclear density distributions ($R_\rho$) were related as:
\begin{equation}
R_V=R_\rho+0.55.
\label{rvrho}
\end{equation}
It was further found that this phenomenological result could be
 reproduced if one used
{\it {density-dependent}} nucleon-nucleon interaction and 
included the exchange part of the
interaction ~\cite{sgh}.
 By this time, a series of papers explored the projectile mass-dependence 
of the optical potentials obtained from phenomenological calculations 
and reported the
following additional empirical observations:
\begin{enumerate}
\item      The equivalent sharp radius for the optical potential
 differed by a fixed amount from the equivalent sharp radius for 
the density (Eq.\ref{rvrho}).
\item  The difference  between the  mean square radii of the potential
 and the density increased with $A$, the target mass, and,
\item The volume integral of the optical model potential (real part) 
per nucleon decreased with $A$.
\end{enumerate}

These observations were also found to be valid for composite projectiles 
like deuterons, $^{3}$He, $^{3}$H, and $^{4}$He (Ref.\cite{msg}).

 The  real part of the (single-) folding optical model potential is written as:
\begin{equation}
 V( r )=   \int d\mathbf{r^\prime} \, \rho(\mathbf{r^\prime})
v(\mathbf{r}-\mathbf{r^\prime}) d\mathbf{r^\prime},
\label{sfold}
\end{equation}
where the density-dependent projectile-nucleon interaction is given by:
\begin{equation}
v(\mathbf{r}-\mathbf{r^\prime})=v_0\left [1-\beta \rho^{2/3}\left(
\frac{\mathbf{r}+\mathbf{r^\prime}}{2}\right)\right ]
f(|\mathbf{r}-\mathbf{r^\prime}|)
\label{vdd}
\end{equation}

Srivastava~\cite{s1} derived analytical relations from
this, when the general form of the density-distribution is taken as:
\begin{equation}
\rho(r)=\frac{\rho_0}{\left[1+\exp\left(\frac{r-C}{a}\right)\right]},
\label{fm}
\end{equation}
where $C\sim A^{1/3}$ is the half value radius,
 and $a$ is the diffuseness of the
distribution. Thus he showed that for short range nucleon-nucleon interaction:
\begin{equation}
R_V=R_\rho+A_1 a \alpha
\end{equation}
and 
\begin{equation}
\langle r^2\rangle_V=\langle r^2 \rangle_\rho+
 \langle r^2\rangle_v+\frac{6}{5}A_1 C a \alpha
\end{equation}
where $A_1$=0.759, $\alpha=\beta \rho_0^{2/3}/[1-\beta \rho_0^{2/3}]$,
and the rest of the terms have the usual meanings.

 It is abundantly clear that the geometrical relations empirically
 observed earlier can be
understood {\it if and only if}, the density distribution of nuclei is
 diffuse (i.e., $a$ is non-zero)
{\it and} the nuclear potential is density dependent
 (i.e., $\beta$ is non-zero).
 If either of these
conditions is not met the geometries of the potentials can't be right
 and the elastic
scattering data can never be explained satisfactorily.
 It was also seen\cite{sgb}
that with these conditions
the volume integral of the potential per nucleon decreased with A.
 These papers appeared
almost ten years after the pioneering papers of Sinha,
 where saturating forces were used for the first time.

These results were then inverted to obtain the range of nucleon-nucleon 
force and the
density dependence of nuclear forces\cite{sbg} and also to propose a
 factorized density dependence
for double-folding models for composite particles~\cite{akc}. 
These relations, along with a generalization of
the Banerjee's theorem to a sum of three vectors, and Satchler's 
theorem for deformed folded potentials\cite{s2}
 were used to derive the
"Srivastava and  Rebel Procedure" \cite{sr1}.
It was used to determine the deformation of density distribution of 
nuclei, when
the deformation of the nuclear potential was known 
from inelastic scattering measurements, in a 
projectile-independent manner. It
was further used to determine the dynamic density dependence 
of nuclear forces as
experienced by projectiles getting inelastically scattered from 
vibrational nuclei\cite{sr2}.

It might interest the reader to know that all this became possible 
because of an analytical evaluation of integrals:
\begin{equation}
I(\nu,q)=\int \, dr \, \frac{r^\nu}{\left[1+\exp
 \left(\frac{r-C}{a}\right) \right]^\nu}
\end{equation}
for arbitrary $q$ and $\nu$ (Ref.\cite{s3}).

It is a very small and yet a very potent example of how a small step can
 open up large
opportunities. It is also important to recall, that in physics, 
developments are often a
function of time and sure enough, several groups often meet a need
 of the hour independently. This integral was evaluated
independently by two French groups, and another group in India, at around the 
same time and is now used extensively in several studies involving 
Thomas Fermi approximation, equation of state, and nuclear matter etc.

The tradition of exploring the consequences of density dependence of 
nuclear forces has been nurtured and kept alive by Dr. D. N. Basu, 
who has now extended it to derive mass
formulae and compressibility of nuclear matter, and to study the
 scattering of unstable nuclei.

\subsection{Optical Model Potential of Loosely Bound Composite Particles}

The target mass and energy dependence of mass-3 projectiles was exhaustively
studied in a work of the group\cite{msg} as mentioned earlier.

The   optical model potential of a loosely bound composite particles has a
distinguishing feature, which  concerns the energy dependence of the
 depth of the real part of the potential. A composite particle may
 dissociate in the field of the target
nucleus  and  later recombine,  giving rise to a non-locality in
 time in the effective one- body interaction. Such a  processes 
 can  contribute  to  the energy  dependence  in  the
optical model potential of composite particles. This aspect was 
 investigated  by Santanu
Pal~\cite{p1} as mentioned earlier in  a  coupled reaction 
 channel  formalism and  it  was  found  that  the
stripping    channel    contributes    about    30\% 
   of    the empirically observed energy dependence of the 
real part of d- $^{40}$Ca optical model potential.
 
Several years later it was seen that Coulomb break up of loosely
bound projectiles like deuteron, lead to a long range dynamic
polarization potential, which has both a real as well as an 
imaginary part~\cite{sr_d}.

Starting  in  the  early 70's, experiments on stripping reactions
 were  extended  to  an interesting domain where particle-unbound
 resonant  states  were populated through stripping
reactions. The measured  stripping  cross  sections
  with  respect  to the total neutron scattering
cross sections displayed  a  strong dependence on the
 orbital angular momentum of the
transferred neutron. In  a theoretical  analysis 
 performed  at VECC, the DWBA formalism was
extended  to  calculate,  without  any  adjustable
  parameter,  the stripping cross sections to unbound states
\cite{p2}
 and succeeded in reproducing the above feature of experimental data. 
The DWBA calculation  was thus  shown  to be  capable of spin
 and parity assignment of the resonant states. An interesting
 feature of the above  calculation was  the  computation  of 
 the  T-matrix  elements involving only scattering
states which have no natural cut-off and thus require
 numerical integration over  the  entire
configuration  space.  A numerical  technique employing
 a convergence factor was developed at
VECC and was shown to handle correctly such slowly converging 
integrals~\cite{p3}. Several years later, analytical results were obtained
for these integrals by Srivastava\cite{s0}.

\subsection{Quantum    Mechanical    Formulation   of   Nuclear Dissipation}

With  the  discovery  of  deep  inelastic collisions in heavy ion induced 
 reactions  above
the  Coulomb  barrier,  it  was almost immediately  recognized  that
  `dissipation'  is  a
fundamental property  of nuclear dynamics in bulk. This discovery
 triggered a great surge of
theoretical  activity in order to understand the origin and nature
 of nuclear dissipation. It was
soon  realized that  a  first-order  theory based on non-equilibrium
 statistical mechanics  would  be
a  good starting point to formulate nuclear dissipation. 
A theoretical model based on linear
response  theory was  developed   in order to
 calculate one- and two-body dissipation~\cite{p4}.
Dissipation essentially portrays a time-correlation
among  T-matrix  elements and  a distinguishing feature
 of the above work was the treatment of
time-correlation arising out of   the  bulk  motion
  which was treated exactly, without invoking any
approximation.
This work showed that one-body friction is much stronger than
the two-body one. The magnitude of one-body friction coefficient
was found to compare favourably with experimental values.
Since incoherent particle-hole excitation  is  the  basic process 
leading  to  dissipation,
particles  can  be  lifted  to  higher single-particle 
  energies   at   higher   bombarding   energies.
Consequently, the radial extent of the relevant T-matrices 
 would be  larger  at   higher
bombarding  energies. This aspect was demonstrated in a
 calculation where it was
shown that the form  factor of nuclear friction depends
 on the incident energy~\cite{p5}.
In order to obtain a    further    insight    into    
dissipation phenomenon in
nucleus-nucleus  collisions,  a  model  theoretical study 
 of two colliding Fermi gases was carried
out~\cite{p6}. 
It was observed that  the  memory  time
for  the  two-body dissipation is significantly smaller than that of
 one-body  dissipation.  A
threshold-type  dependence  of  the transferred  energy 
 on  the  relative  velocity  between the two
nuclei was also observed. It was further observed that the 
 total dissipated energy due to one-body
processes is shared between the two  nuclei  approximately 
in the ratio of their masses. The rate
of energy transfer due to one-body dissipation was also found  to be close
 to those derived from experimental data.

Angular  momentum is also dissipated along with kinetic energy 
in deep inelastic heavy ion
collisions.  A  theoretical  model  for angular  momentum 
 transfer based on one-body perturbation
theory was   also  developed  at  VECC~\cite{p7}.
 Comparison of
theoretically calculated values  with  experimental results
  on magnitude of angular momentum
transfer and its degree of misalignment established that 
 inelastic  excitations  are  as important
as  nucleon  exchange  processes  in producing angular momentum dissipation.

A  significant  step  was  taken  when  an  exact  calculation of 
one-body nuclear dissipation was performed without  invoking 
 the first-order    perturbation    approximations~\cite{p8}.
    Time-dependent       antisymmetrized single-particle
wave  functions  were calculated for a colliding nucleus-nucleus  
system  from  which   the
velocity   dependent dissipative  force  was  subsequently  extracted.
 The strength of this
theoretical dissipation was found to  lie  between  the  two 
phenomenological  models  prevailing
at  that  time,  namely the surface and  proximity  frictions.  
It  was  further  shown  from theoretical
calculations  that  nucleon exchange can account for about
 1/2 to 1/3 of total one-body  friction,
the  rest  of  the strength being due to inelastic excitations.

The theoretical  tools     developed     for    time  
  dependent single-particle wave  functions
were  subsequently  extended  to calculate the absorptive part of 
 nucleus-nucleus  optical  model
potential.  In  a  model  calculation  performed  at  VECC~\cite{p9},
it  was
shown  that  nucleon  transfer contributes  
about  half of the phenomenological strengths of the
absorptive potential.

\subsection{Break-up of Light Ions in the Nuclear and Coulomb Field of Nuclei}

Break-up of light ions like deuterons, $^{3}$He, $^{3}$H, $^{4}$He, $^{6}$Li, 
etc. constitute a large part of the total reaction cross-section, and carry
information about the relative motion wave-function of projectile fragments. 
These processes become the largest contributors to the cross-sections
for loosely bound particles, e.g., radio-active nuclei like $^{11}$Li etc.

With this in mind a prior-form distorted wave Born approximation~\cite{sr_p}
treatment
was developed, which was most suitable for studying Coulomb break up
\cite{feature} as well
as break-up of composite projectiles like $^{6}$Li. A series of papers
studied various features of these processes and led to formulation of a
procedure to try to extract cross-sections of astro-physical 
interest~\cite{astro}.
These developments led to confirmation of interference effects~\cite{inter}
among the projectile fragments as well as to identification of orbital 
dispersion on them\cite{orbit}, for the first time in the literature at 
energies of about 25 MeV/N.

The prior form distorted waves Born approximation developed for this 
purpose was used to get a direct measurement of off-shell T-matrix for
projectile-target interaction,
again for the first time in the literature\cite{off}.

Dr. D. N. Basu has continued to contribute extensively to this field and
has helped plan several detailed experiments at the pelletrons at Bombay and
New Delhi. 

\section{Nuclear Theory Group Comes of Age}

By this time, the theory group of the VECC had several practitioners and
over years they made very valuable contributions. We shall briefly recall
the major initiatives in the following.

\subsection{Deep Inelastic Collisions}

       The Deep Inelastic Collisions (DIC) between heavy ions are
characterized with a
lot of energy loss and angular momentum from the relative motion
 to the reacting
fragments. Naively, the angular momentum gained by the reactants
 should be oriented
normal to the reaction plane. However, in reality, there is
sizable dispersion in their
orientation and they are normal to the reaction plane, only on
the average. 

The energy
and angular momentum damping are generated through the
nucleon exchange mechanism
as long as the densities of the reactants overlap.
The nucleon exchange between the
nuclei is a random process. Moreover, the intrinsic
 Fermi velocities of the nucleons,
which get added to relative velocity, are randomly distributed.
 This random motion
would give rise  to random component of the transferred angular momentum.
 Some simple estimates of angular momentum misalignment, based on this
 conjecture, without
taking into consideration the dynamical nature of the process and
the quantal nature of
the nucleon exchange mechanism, were available in the literature
and could qualitatively
explain the observed data for peripheral collisions, only.

 However, we followed the same
idea and made detailed dynamical calculations and
could explain the observed data
quantitatively, for peripheral as well as DIC processes.
It was observed that the quantal
nature of the nucleon exchange mechanism,
 i.e. Pauli correlation, is vital to explain the
experimental angular momentum misalignment data~\cite{kk1}.

        Further detailed studies were made to understand
spin dispersion and alignment in
DIC in the frame work above mentioned stochastic nucleon exchange model
incorporating, explicitly, the temperature dependent
intrinsic Fermi velocity distributions,
the inter-nuclear barrier and the shell gap in the single particle spectra.
 The temperature
dependent Fermi velocity distributions increase the available phase space
for nucleon
transfer from one nucleus to other whereas the shell gap
 reduces it and the interplay of
these two affect the physical observables  quite significantly.
These
detailed dynamical calculations were quite successful in explaining the
experimental data~\cite{kk2}.

        In early 1980's, it was still an open question as
 to how the energy damped in DIC
processes from relative motion is partitioned
between the two reactants. The calculations
pertaining to the various physical observables of
deep inelastic collisions were either
performed in the zero temperature limit or with
equilibration of energy. 
Detailed dynamical calculations were performed in the stochastic
nucleon exchange model to study the
 evolution of the excitation energies of the reactants.
 It had been observed that for
asymmetric systems, the energy is shared equally between
 the fragments for peripheral
collisions, i.e. low energy loss and the system approaches
 to-wards equilibration for very
deep collisions, i.e. large energy loss\cite{kk3}
 The time
evolution of the temperature of the reactants at a
 given impact parameter for a typical
reaction was also obtained.

\subsection{Heavy Ion Reactions in Fermi Energy domain}

\subsubsection{Promptly emitted particles}

In heavy ion reactions with energies well above
the Coulomb barrier and in the
Fermi energy domain, a long tail is observed in the particle
energy spectra; a
consequence of the particle emission in the early stages
of the reaction. These energetic
particles are called the promptly emitted particles (PEPs)
 or Fermi jets. These particles
carry away linear momentum and energy from the system, and
 in fusion-like reactions
one is left with incompletely fused systems with
 excitation energies and linear momenta
less than those of corresponding compound nuclear systems.
 Based on the nucleon
exchange mechanism, the promptly
 emitted particle (PEP)
model was extensively 
developed to study the heavy ion reactions in the Fermi energy
 domain. Basic essence of this
model is the emission of 1-body or primary PEPs and
 2-body or secondary PEPs. The
relative velocity of the transferred nucleon is
 boosted by its coupling with the intrinsic
Fermi velocity. However, a part of the transferred flux,
 while passing through the
medium of the recipient, may be completely absorbed due to
 collisions and the rest may
be emitted into the continuum, provided the energy is
 sufficient to overcome the nuclear
barrier. These particles which have suffered no collisions
 along their path are called 1-
body PEPs or primary PEPs. The absorption may, however,
be reduced because after the
first collision suffered by the transferred nucleon and
depending upon their energies, any
one of the colliding nucleon or both of them may be
 emitted in the continuum. These
emitted particles are called 2-PEPs or
 secondary PEPs.
Through extensive work using  the PEP model it has been
 shown explicitly that the
inclusion of secondary PEPs in the calculations is
very crucial in explaining:
\begin{enumerate}
\item The high
energy tails in the experimental particle spectra
for both neutrons and protons for a wide
range of systems and incident energies\cite{kk4}.

\item The observed
 saturation of the linear momentum
transfer per incident nucleon ( $P_T/A$) and of
temperature or energy deposited  in the
system in the intermediate energy domain.
It had been observed that more the incident
energy, more PEPs are emitted from the system leading
 to the saturation of linear
momentum transfer and energy deposited\cite{kk5}.

\end{enumerate}

\subsubsection{Incomplete fusion}

        Experimentally it had been observed that in
heavy ion reactions with energies
above $\sim$10 MeV/A, the residual velocity ( $V_R$ ) of
the fused system is larger ( smaller) as
compared to compound nuclear velocity ( $V_{\mathrm{CN}}$) for
inverse kinematical ( direct) reactions
and that for symmetric systems $V_R= V_{\mathrm{CN}}$.
 Qualitatively, these observations were explained
with the assumption that smaller of the two reactants
 loses more particles in the initial
stages of the reaction. We made calculations for the
 residual velocity in the frame work
of the PEP model, with inclusion of 2-body PEPs.
Our results are in conformity with the
experimental data\cite{kk6}.

 With the increase in incident
energy the importance of the
loss of nucleons from the pre-equilibrium phase increases\cite{kk7};
this was very clearly demonstrated in a study of the time
evolution of the loss of the particles from the reactants.
The study also demonstrated that it is indeed the smaller
 partner which loses more particle in
the reaction.

        In the intermediate energy heavy ion fusion-like
reactions, the whole reaction
scenario can be thought as  of comprising of two phases:
An initial pre-equilibrium phase
where a number of energetic particles are emitted which
 carry away energy, linear
momentum and angular momentum leading to an
 incompletely fused composite (IFC),
and the second phase in which the highly excited
 composite de-excites through statistical
processes yielding evaporated light particles and
final residues.

 In order to understand
this  reaction scenario we made a fully dynamical
 calculation entailing the evolution of
the nucleus-nucleus collision process where the
 pre-equilibrium phase is followed by the
de-excitation of  IFC through binary sequential
 decay process, on event-by event basis
using Monte-Carlo simulation technique.
 There is no free parameter involved in the
whole calculation, from the initial contact point
 of the colliding nuclei to the final
residues, and the calculations provide an
accurate description of the experimental data\cite{kk7}.

        The non-fusion processes, especially, the deep
 inelastic processes in intermediate
energy heavy ion collisions, like fusion processes,
are also `incomplete' in the sense that
the excitation energies deposited in the two fragments
 are significantly smaller than the
total kinetic energy loss from the entrance channel.
 These processes where a fraction of
entrance channel kinetic energy is carried away by the
pre-equilibrium emission, thereby
reducing the fragment excitation energy are called
`Incompletely Deep Inelastic
Collisions' (INDIC) processes. 

We developed an integrated
 theoretical model in which
the dynamical evolution of the colliding system,
in the Fermi energy domain, leads to the
formation of either IFC or incomplete deep inelastic
 and-or quasi-elastic fragments,
which subsequently undergo statistical binary decay to
 yield the final residues. This
model was used to study the salient features of INDIC
and incomplete fusion, in detail
and was applied to calculate the intermediate mass fragment
 (IMF ; $3\leq Z \leq 25$) yields, and was found to be very
successful indeed\cite{kk8}.

 It  was also found that
 that IMF with $Z< Z$(projectile) are almost entirely
 emitted through INDIC
processes.

\subsection{Fission}

        The experimental observation of pre-scission neutrons
indicated that the
phenomenon of fission is a dissipative process where
 the shape degrees of freedom or
collective degrees of freedom during their evolution
 interact with the nucleonic degree of
freedom and dump their initial kinetic energies
in to the system as excitation energy,
causing the emission of pre-scission neutrons.
In a sense, thus, the dynamics of fission
process resembles the standard `Brownian motion' problem.
The collective degree of
freedom called the `Brownian' particle interacts
stochastically with the nucleonic degree
of freedom, called the `surrounding bath' and
 dissipation  is generated through their
mutual interaction.

In order to understand the dynamics of
 fission process, we developed a model,
first, by taking a simple shape for the fissioning
nucleus i.e. two leptodermous spheres
connected by a neck. This shape, introduced by
Swiatecki, reduces the collective degrees
of freedom to only one- the surface to surface
separation. In this model the system is
initially placed in the minimum of the potential
with a fraction of the initially available
energy assigned to the collective variable.
 This is done with the assumption of equal
probability of the system being in any microstate
 and is realized by using a uniform
random number distribution. The radial and tangential
 friction coefficient are calculated
using Werner-Wheeler method, assuming the
 system to be irrotational  hydrodynamical
fluid. This model was successfully applied
 to calculate the IMF yields and total kinetic
energy of the fragments for fusion-fission systems
 below the Businaro-Gallone point
where the asymmetric fission dominates the
 symmetric fission\cite{kk9}

        This simple dynamical model of fission was
further used to calculate  pre-scission
neutron multiplicity. A systematic study of
 relationship between pre-scission neutron
multiplicity and nuclear viscosity was done
 for wide range of mass (150--200) and incident
energy (4--13 MeV/A). The values of the viscosity coefficients,
which were used to
predict the experimental pre-scission multiplicity,
were found to follow a global relation\cite{kk10}:
\begin{equation}
\mu (E/A, A_{\mathrm{CN}})=a\frac{E}{A}+b A^3_{\mathrm{CN}}.
\end{equation}
where
\begin{equation}
a= 0.160 \pm 0.023,
\end{equation}
and
\begin{equation}
b=0.357\times 10^{-6} \pm 0.26 \times 10^{-7}.
\end{equation}

        The fission dynamics had been studied earlier
 either by solving the Langevin
Equation or multidimensional Fokker-Planck Equation,
which is differential version of
Langevin Equation. However, we developed an alternative
 approach based on the fact
that for stochastic processes, the full solution
of Fokker-Planck equation admits an
asymptotic expansion in terms of the fluctuations,
provided variances of the physical
observables are small compared to their mean values.
This approach had been used
earlier by Van-Kampen for stochastic processes
 with a constant value of diffusion constant.
We generalized the asymptotic expansion method in
the case of fission because then
the dissipation depends on the instantaneous shape
 of fissioning system and thereby the
diffusion constants are shape dependents.
The expansion, in its zeroth order, yields Euler-
Lagrange equations for deterministic motion and
in its first order one gets equations for
calculating the accompanied fluctuations.
To the best of our knowledge, such an
approach in case of fission is not available in the literature.
As compared to our earlier work, here we used a
 generalized realistic shape
parameterization for the shape of the fissioning
nucleus and corresponding shape
dependent two-body and one-body friction coefficients
were calculated using the Werner-
Wheeler method, as earlier. With a single value of
 viscosity coefficient we could
reproduce the experimentally observed total kinetic
 energy and it variance,
 pre-scission neutron
multiplicity for both  symmetric and
asymmetric fission and pre-scission  neutron
energy spectra\cite{kk11}.

\subsection{Limiting Temperature in Nuclei and Nuclear Equation of State}

The thermostatic and thermodynamic properties of infinite and
finite nuclear systems is another key area of interest where
we are intensely pursuing research for over a decade. We
showed how a 'limiting temperature' for finite nuclei follows
naturally from thermodynamic analysis \cite{ban}. With a
chosen effective nucleon-nucleon interaction, the nuclear
equation of state (EOS) follows directly. With explicit introduction of
spin degrees of freedom, the nuclear EOS was employed to understand
the properties of neutron stars; particularly interesting is the
ferromagnetic phase transition in neutron stars at a density
$\sim$3-4 times the density of normal nuclear matter \cite{uma1}
and the possibility of a phase transition from baryonic matter
to quark matter at higher densities \cite{uma2} in the core of
the neutron stars.

The nuclear EOS has also an extremely important bearing on understanding
the fragmentation of nuclei to pieces (nuclear multi-fragmentation)
and its possible relationship to the nuclear liquid-gas phase transition.
The statistical multi-fragmentation model is normally accepted as the
standard model for nuclear disassembly; it has been used to explain
the mass or charge distribution in nuclear fragmentation and has
been of enormous importance in calculating the temperature of the
fragmenting system (from double-isotope ratio) and then draw
inferences about the liquid-gas phase transition in nuclei. Our
calculations showed for the first time \cite{pal} that such
inferences are ambiguous. After disassembly there is enough phase space
for the fragments to recombine thus appreciably changing the scenario
of the fragment production and the associated inferences. On the
other hand, from a full microscopic calculation of the EOS of
finite nuclei \cite{de3} in a finite temperature Thomas-Fermi
(FTTF) framework, we are able to show that finite nuclei exhibit
signatures of liquid-gas phase transition \cite{de4,tap1} at
temperatures far below the critical temperature for infinite
nuclear systems. We further showed that in the preparation of hot
nuclei, the compressional degrees of freedom enhance the liquid-gas
phase transition signatures considerably \cite{sam1}.

We have further studied the stability of nuclei beyond
the drip lines \cite{tap2} in the presence of an enveloping gas
of nucleons and electrons as prevailing in the inner crust of
a neutron star in the FTTF framework. We predict a 'limiting
asymmetry' in the isospin space beyond which nuclei can not
exist even in the stellar matter. The ambient conditions such as
temperature, baryon density and neutrino concentration in which
the nucleation process of the different species of these
exotic nuclear clusters occurs from the nucleonic sea as the
neutron star cools down in the early stages of its formation
have also been studied in detail.

\subsection{Nuclear Structure}

        It has been observed experimentally that the
 charge radii of Ca-isotopes increase
with the addition of neutrons up to the first
 half of the $ 1f_{7/2}$ shell and then decrease in
such a way that after filling the shell the charge
 radii of $^{40}$Ca and $^{48}$Ca are almost equal. It
is well known that ground state neutron-proton
 correlations are solely responsible for
such a modulation but it is not known how this
 n-p correlation affects this observed
isotopic shifts. These n-p correlations,
 however, are manifested through the occupancies
of the single particle levels. Thus, taking into
consideration the experimentally observed
single particle occupancies, which reflect n-p
correlations, we have been successful in
reproducing the parabolic behaviour of charge
radii of Ca-isotopes using a one body
Woods-Saxon type potential. The parameters of
 this potential were fixed by reproducing
the single particle energies and Fourier-Bessel
 coefficients for the charge distribution of
$^{48}$Ca nucleus. It may be mentioned that keeping all
 the other parameters fixed, the
diffuseness parameter $a$ had to be modified by $\sim 10\%$
for
$^{40}$Ca - $^{46}$Ca. This may be
indicative of neutron skin effect as one goes on
 adding more and more neutrons~\cite{kk12}.

        One neutron halo nucleus $^{11}$Be, with  the last neutron
 separation energy equal to 0.5
MeV, has been observed to have peculiar character of parity
 inversion of its ground state
and first excited state. This long standing problem has been
 successfully solved by us by
using particle-vibration coupling model and the predicted low
lying energy levels and
spectroscopic factors agree very well with the experimentally
 observed ones. The single
particle occupancies, used in the model calculations,
 were taken  from Hartree-Fock
calculations. The only free parameters in the calculations
were the relative energy
spacings of $2s_{1/2}$  and $1d_{5/2}$ single particle orbitals with
 respect to $1p_{1/2}$ orbital and the
coupling strength. This success of this model was further
 tested in the case of one neutron
halo nucleus $^{19}$O. The predicted low lying energy spectra
and the spectroscopic factors
had an excellent agreement with the experimentally observed
 ones\cite{kk13}.

\subsection{Quantum Chaos and Nuclear Dynamics}

From  the  middle  of 1980's, chaos in quantum mechanical
systems became a topic of
intense research in order to answer a number of profound
 theoretical questions such as the
signatures of quantum chaos and its correspondence with
classical  dynamics.  Though  a
compound  nucleus  with  its  numerous resonance
 states was early recognized as a benchmark
for quantum chaos, it required a series of detailed
 investigations  to  establish  the  chaotic nature
of nuclear   dynamics   in   other   degrees   of
   freedom    (e.g. single-particle  or  collective).  The
spectral properties of the single-particle states  in
  a  two-centre  potential  model  were studied  at
VECC  in order to measure the chaotic content of the
 system\cite{p10}.

 A regular-chaos-regular  transition  was observed as the two-centre
potential was evolved to represent the
 approaching phase  of  two heavy  nuclei.
Subsequently this work was extended by including
spin-orbit potential and the  detailed  nature  of
the wave   functions   was   studied~\cite{p11}
 in
order  to  distinguish  regular  and irregular  states
 in  the system. 

The spectral fluctuations in a
two-centre shell model potential with spin-orbit
 interaction were next analyzed in detail and it was
demonstrated that a  effective underlying classical
 dynamics can be identified though spin-orbit
potential  has  no  obvious  classical  analogue\cite{p12}.

Work  at VECC on quantum mechanical systems without any classical
analogue continued
further by considering the Dirac  equation.
 A Dirac  particle  was considered in cavities of various
shapes and numerical evidence was obtained  of  the
 influence  of  periodic orbits  in  the  quantal
density of states\cite{p13}
 and  interestingly,  the
orbit lengths turned out to be  same  as  that  of
spin-less  case  though  the associated  phases
were  different.  In  a detailed study of the wave-functions,
 existence of scarred states  and
also contour splitting of irregular wave functions,
typical features  of  systems  with  a
classical analogue, were observed which demonstrated
 the  existence  of  an  underlying
classical dynamics for the Dirac Hamiltonian.

\subsection{Chaos and Dissipation}

One  of the main motivations to study chaotic features in
 nuclear dynamics is the fact that
the response of a system depends on the nature  of  its
 intrinsic motion. With this view in mind,
chaotic dynamics of single particles in axially symmetric
 nuclear  shape was      investigated    at
VECC~\cite{p14}.

 Shapes of different multi-poles
were considered in this study and the  systematic
 dependence  of the degree of chaos on
deformation parameters was extracted. Such studies
 provided  the background for an important
development in dissipation theory crafted at VECC.
 It was shown  on  theoretical grounds that the
one-body wall friction should be modified taking 
into account the degree of  chaos  in the intrinsic
dynamics of a nucleus\cite{p15}.  

A new friction namely  the  Chaos  Weighted Wall  Friction  (CWWF)  was
formulated  and used successfully in subsequent applications. 
 From a  model  study  of  an  ideal
gas undergoing  volume  conserving  shape  oscillations,
it was found that the chaos weighted wall
friction provided a fairly  reliable picture   of
  one   body   dissipation\cite{p16}.
Subsequently the chaos weighted  friction
was  applied  to  the  surface  motion  of  a
cavity  undergoing fission-like shape evolutions and the
energy damping was found  to compare favourably with the
 irreversible energy transfer obtained
from  an  exact  dynamical  calculation\cite{p17}.

\subsection{Langevin Dynamics of Nuclear Fission}

\begin{figure}
\centerline{\psfig{figure=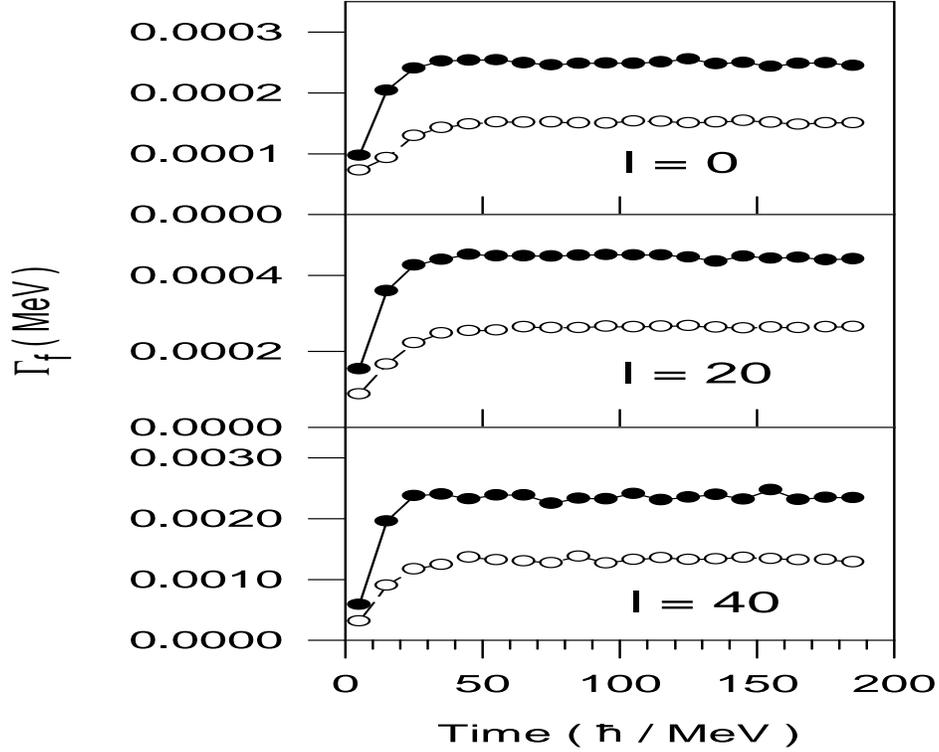,height=10cm,width=12.5cm}}
\caption{
Time-dependent fission widths calculated with chaos weighted wall
friction (solid circles) and usual wall friction (open circles) for
different compound nuclear spins $L$}
\label{fig_pal1}
\end{figure}

Experimental studies   during   the   last   decade  or
  so  have established that the
statistical model of  Bohr  and  Wheeler  is inadequate
  to  describe fission of heavy compound
nuclei at high excitations. The need for a dissipative
 force in dynamical  model calculations  was
recognized soon, the strength of which however was
 found to be much lower than the established
model of one-body wall friction. Since  a  suppression
  in  the  strength  of  wall friction  was
obtained at VECC earlier through the formulation of
 chaos weighted wall friction, a detailed
program  was  undertaken to  perform  fission dynamics
calculations using this friction in the
Langevin  equations.  The time-dependent fission widths
were first calculated in order to find their
dependence on strength of dissipation.  The  fission  widths
  calculated  at VECC using the chaos
weighted wall formula  were  found  to be larger by about
 a factor of 2  compared  to  that
obtained  with  the  usual  wall friction (Fig.\ref{fig_pal1})~\cite{p18}.

\begin{figure}
\centerline{\psfig{figure=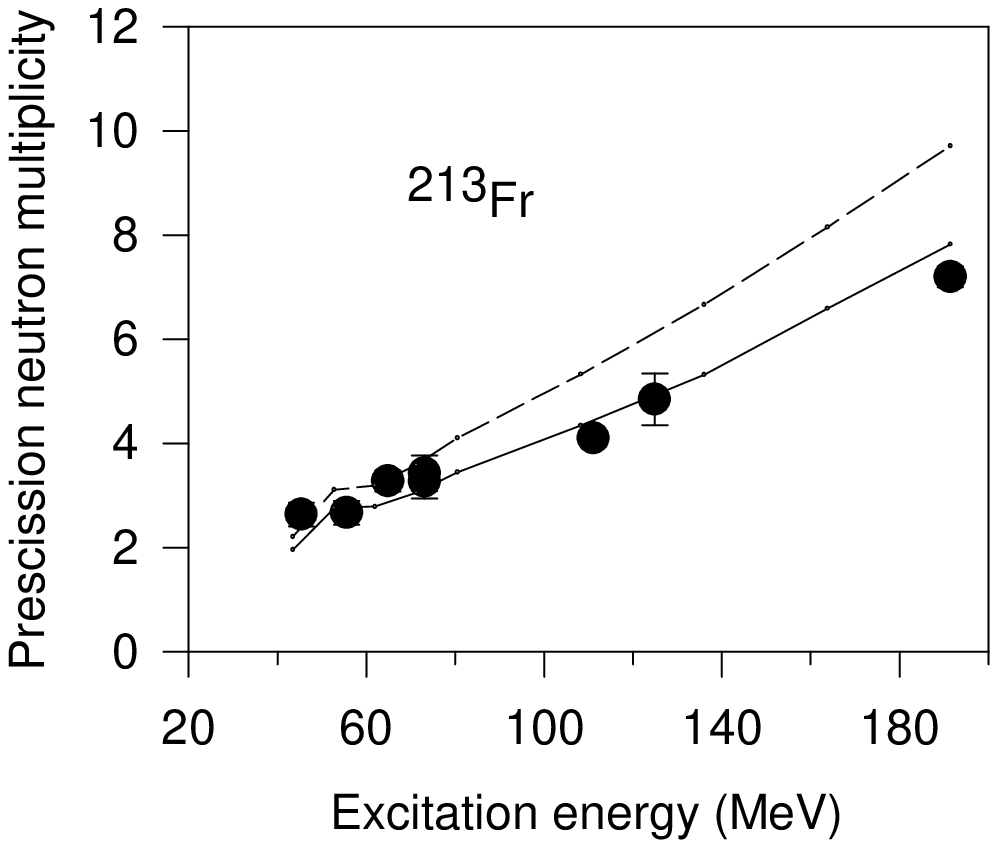,height=10cm,width=12.5cm}}
\caption{
Calculated neutron multiplicities with chaos weighted wall friction
(solid line) and usual wall friction (dashed line) along
with the experimental values.} 
\label{fig_pal2}
\end{figure}

It was  further  observed  from  the calculated  values  of  the
time-dependent fission widths that a steady flow
 to-wards the scission point is established, after
the initial  transients,  not  only  for  nuclei
  which  have fission barriers but also  for  nuclei  which
have  no  fission  barrier\cite{p19}.
 Subsequently, the
statistical emission of neutrons  and  photons
 were coupled with the dynamics of fission in the
Langevin equations and pre-scission neutron
multiplicities and fission probabilities were  calculated
for  a  number  of  systems \cite{p20} (Fig.\ref{fig_pal2}).

A detailed analysis of  our  results led  us
 to  conclude  that  the chaos-weighted wall friction can
adequately describe the fission dynamics in the presaddle region.

\begin{figure}
\centerline{\psfig{figure=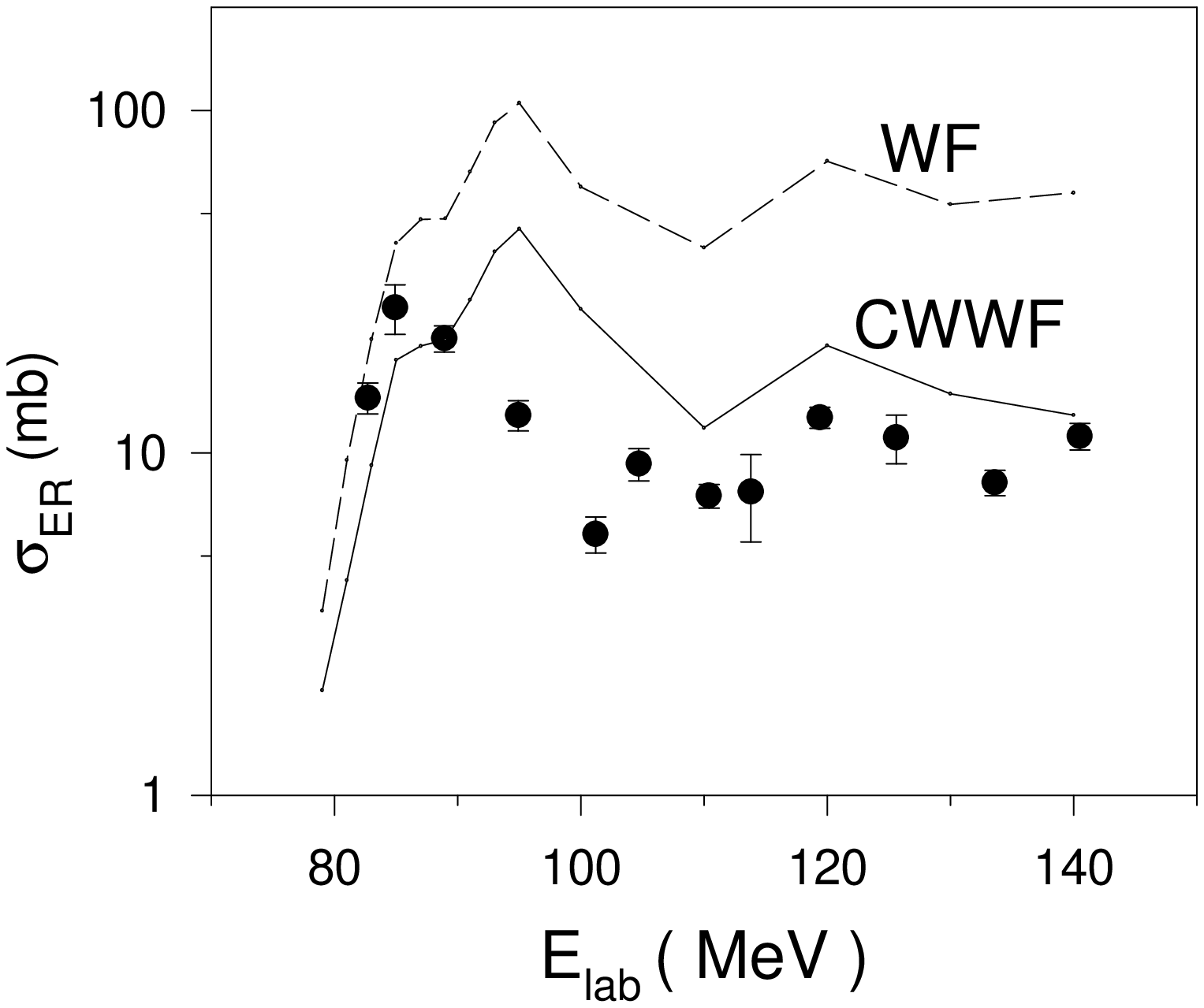,height=10cm,width=12.5cm}}
\caption{
Evaporation residue cross-sections calculated with chaos 
weighted wall friction
(solid line) and usual wall friction (dashed line)  for $^{16}$O+$^{208}$Pb
along
with the experimental values.} 
\label{fig_pal3}
\end{figure}

Evaporation residue cross-sections were next
  calculated at  VECC in   the   framework   of
the  Langevin  equation  coupled  with statistical
 evaporation of light particles  and  GDR's\cite{p21}.
 The  evaporation residue cross-section was
found  to  be  very  sensitive  to  the choice of
 nuclear friction (Fig.\ref{fig_pal3}).
 The results indicated that the
chaotic  nature  of  the single-particle dynamics
 within the nuclear volume can provide an
explanation for  the  strong  shape dependence  of
 nuclear  friction which is usually required to fit
experimental data.

\subsection{Unified Description of Nuclear Matter, Scattering, and
Radioactivities}

In an exhaustive study,
a realistic density dependent effective interaction has been  used to
calculate nuclear incompressibility, proton, alpha and cluster
radioactivities,  elastic and inelastic scattering cross-sections,
and nuclear masses.

\begin{enumerate}
\item The microscopic nucleon-nucleus interaction potentials are obtained by
folding the density dependent M3Y effective interaction supplemented with
a zero-range pseudo-potential  to account for the exchange
term, with the density
distribution of the nucleus. 

\item The microscopic nucleus-nucleus
interaction potentials are obtained by double folding the same
interaction along with a factorized density dependence term~\cite{sbg} 
to account for the saturating properties of nucleus
with the densities of the nuclei.

\item
 The density
dependence parameters of the interaction have been obtained
from nuclear
matter calculations, which give a
 reasonable value for nuclear incompressibility\cite{Ba04}.

\item The quantum
mechanical tunneling probabilities for nuclear decays are calculated
within the WKB approximation, which provides the
lifetimes for proton, alpha~\cite{Ba03},
 and cluster radioactivities~\cite{Ba02,BA03} in
good agreement with the experimental results over a wide range spanning
about thirty-five orders of magnitude. The life-times of the alpha
decay chains of the recently measured super-heavy nucleus having
Z=115, has also been estimated accurately~\cite{BA04}.

\item The same nuclear interaction
potential when used as the optical potential provides a good description for
the elastic and inelastic scattering of protons\cite{Gu05}.

\item The parameters of the density dependent nucleon-nucleon interaction
are used to get an accurate value for  the mean free path of nucleons
in nuclear matter\cite{Gu05}.

\item And finally, the 
co-efficients in the Bethe-Weizsa\"cker mass formula are
obtained by fitting\cite{Ch05} the experimental atomic masses and to get
the saturation
energy per nucleon and the
equation of state for nuclear matter.

\end{enumerate}

\section{The Next Frontier: The Quark Gluon Plasma}

While others will talk about the opening of new frontiers on the accelerators
and experimental nuclear physics at VECC under the leadership of Dr. 
Bikash Sinha, we focus
our attention on the studies of relativistic heavy ion collisions and 
signatures of quark-gluon plasma. 

By now it is fully established that quantum chromodynamics (QCD) describes the
strong interaction between quarks and gluons, which constitute the hadrons. 
Let us not forget that QCD is responsible for almost 95\% of the mass of
 hadrons. One of the most spectacular predictions of QCD is that under 
conditions of extreme temperatures or pressures, the quarks and gluons 
which remain confined inside hadrons are de-confined and a novel state of
matter called quark gluon plasma (QGP) is created. It is believed that our 
universe, which started as a Big Bang, was in the state of QGP a few 
micro-seconds after the Big Bang, before it cooled further, and produced
first neutrons and protons, and then galaxies and stars, etc. It is also 
expected that such matter may form the core of neutron stars, where it may
have a low temperature ($~$ 5--10 MeV) but very high baryonic chemical
potential.

The present excitement in the field is due to the expectation that QGP can
be created in relativistic heavy ion collisions. This has led to international
collaborations leading to experiments at CERN SPS, Brookhaven Relativistic Heavy Ion Collider, and CERN Large Hadron Collider (under construction). We have 
already heard from Dr. Y. P. Viyogi, about our participation in these 
experimental ventures.

Dr. Bikash Sinha introduced us to the charm of electromagnetic signatures
of quark gluon plasma. If a QGP is formed, quarks and anti-quarks may 
annihilate to produce photons or dileptons, or quarks and gluons may 
scatter to produce them. Because of their electromagnetic nature, photons
and dileptons interact only weakly with the system and carry the 
information about the conditions of their birth. 

\subsection{The Initiation}

The early calculations estimated the production of photons only from the
QGP phase. It was soon realized~\cite{q1} that photons would be produced 
in the QGP phase as well as in the quark matter in the mixed phase, and the
hadronic matter in the mixed phase and the hadronic phase. This was the
first estimate in the literature for production of photons from the entire
history of the system. It was also realized that hadronic phase was not likely
to be consisting of only pions, as was usually assumed in the literature.
As a first step, the hadronic matter was considered to be consisting of
only the lightest hadrons ($\pi$, $\rho$, $\omega$, and $\eta$) and this 
already led to a considerable reduction in the life-time of the system,
due to increase in the number of degrees of freedom
 in the hadronic phase~\cite{q1}.

The early years of the field of QGP were beset with debates about the so-called
Bjorken and Landau hydrodynamics and concepts of boost-invariant expansion of
plasma. A classic study, starting from first principles was used to clarify
this issue, and to explore the consequences of boost-non-invariance on the
rate of cooling. It was also shown that both the Landau and Bjorken
hydrodynamic solutions emerged naturally from the Telegraph Equation, when 
different boundary conditions were applied~\cite{q2}.

While exploring the Landau's hydrodynamics, we found that the approximations
used to get a Gaussian distribution 
commonly employed, for the multiplicity density are not
at all satisfied! However, they {\it were} satisfied if one could assume
that the speed of sound is very small. This suggested that the dynamics of
the relativistic
heavy ion collisions at AGS and SPS energies could be dominated by a mixed 
phase, while for those at lower energies, a hadronic phase prevailed,
which however had contributions from massive hadrons which 
reduced the speed of sound considerably~\cite{q3}.

Around this time, the first  accurate
estimates for the rate of production of photons
from QGP as well as hadrons were obtained. To every-ones surprise they
came out be almost identical, leading to a debate where photons could
at all distinguish between QGP and a hot hadronic matter. 
It was shown by us that when the dynamics of the 
evolution of the system was taken into account, QGP being at larger
temperature radiated more photons having large transverse momenta, 
compared to the hadronic phase when the temperatures are small~\cite{q4}.
This work, re-established the uniqueness of electromagnetic probes of
the quark-gluon plasma, once for all.

Till this time, all calculations, including our own studies had completely
ignored the transverse expansion of the system, which becomes significant,
if the life-time of the system is large. The resulting radial flow cools the
system rapidly. It  also imparts additional transverse momenta to the emitted
particles, thus mimicking a larger apparent temperature. The first ever 
calculation, incorporating these effects for photons were reported by us,
suggesting large modifications to the windows where photons from the
quark-matter were likely to dominate~\cite{q5}. This procedure has now become
standard in the literature. Some years later, this work was further 
extended~\cite{q6} to describe the hadronic phase as 
consisting of all the hadrons
in the particle data book (having $M <$ 2.5 GeV) and in complete thermal and
chemical equilibrium, a fact supported by the success of thermal models
in explaining the particle ratios measured in the relativistic heavy ion
collisions. it was also shown that the electromagnetic signature of the
plasma were quite sensitive to the equation of state for the hadronic 
matter~\cite{q7}.

Hanbury-Brown Twiss interferometry is a very useful probe for getting 
information about the size of the sources and their evolution. Pion 
interferometry has been used extensively for this purpose. However pions
are mostly emitted at the time of freeze-out
and experience final state interaction. Photons on the other hand
are emitted at every stage of the evolution of the system, and thus it
was suggested by Sinha that photon interferometry could be used to get
information about the early stages of the system. This was investigated
in a series of papers which established the usefulness of photon interferometry
in getting the size of the system during different phases of evolution and
also in getting the life-time of the source\cite{q8}. Several years
later these studies were repeated with improved rates for production of
photons and evolution of the system~\cite{q9}.  Results with  an 
special emphasis
on photons having very low transverse momenta~\cite{q10}
for which the first ever photon intensity interferometry
experiment
for relativistic heavy ion collisions
 has just been conducted have also been obtained. Studies were
also reported using pre-equilibrium photons from a parton cascade
 model~\cite{q11}.

A series of papers studied the importance of soft photons
and very low mass dileptons in chronicling the last stages
of the system formed in such collisions\cite{q12}. These were
found to be a very accurate probes of the flow in the hadronic phase.
The production
of photon pairs was also investigated in detail~\cite{q13}.

\subsection{The Baptization by Fire}

By this time the first results for single photon production in S+Au
collision at CERN SPS were reported. The preliminary data was analyzed by us 
with a startling result: If we assumed that there was no phase transition 
to quark gluon plasma in the collision, then we considerably over-predicted 
the data, while calculations with the assumption of the formation
of quark gluon plasma, which expanded, cooled, entered into a mixed 
phase of QGP and hadrons and then underwent a freeze-out from a 
hadronic phase gave results which were consistent with the 
measurements~\cite{sau_gam}.  This was the first ever indication, involving
thermal photons, that a quark hadron phase transition may have taken 
place in these collisions. This work generated a lot of discussion
and attempts were made to analyze the dilepton mass spectra measured
for the same collisions, using the same model~\cite{sau_dil}.

This however led to an important indication that there has to be
a large modification of the spectral function for hadrons in the hot and dense
hadronic matter. 

An exhaustive and elaborate study was then planned to understand the
rate of production of photons and dileptons from a hot hadronic
matter where medium modification of hadron properties were explicitly
accounted for (see later).

Some years later, the CERN SPS experiments exploring the collision of
lead nuclei again reported the single photon production. These data
also showed that either a quark gluon plasma had been produced in the
collision\cite{ss_pb}, or a massive modification of the hadronic properties 
had taken place (see later). These calculations used photon 
productions from the quark matter up to two-loops and employed 
rich hadronic equation of state. A reanalysis of the earlier 
measurements for S+Au collision was also done- the final data gave
only upper limits of the photon production, but the conclusions remained
unaltered\cite{sau_r}. The same model of evolution was again used to
get the large mass dilepton production in the collision of lead
nuclei, with similar conclusions\cite{pb_dil}.

In the following we briefly recall some of the major initiatives in the
field of QGP taken by the theory group.

\subsection{Successive Equilibration in QGP}

One of the most important quantities required to study the signals and
properties of QGP is its initial  thermalization time.
Suggestions of an early thermalization were given~\cite{LHC}
which admitted large initial temperatures, limited by the
uncertainty principles that $\tau_0 \sim 1/3T_0$, where $\tau_0$
is the initial time and $T_0$ is the initial temperature
of the QGP.

 Within
the framework of Fokker-Planck equation
it has been shown  that the approach
to both kinetic~\cite{preq1} and chemical equilibrium~\cite{preq2}
 in a quark gluon system formed
in the ultra-relativistic
heavy ion collisions proceed through a succession of many time scales,
significantly affecting the signals of QGP formation~\cite{preq3}.

A very detailed calculation of the effect of transverse expansion
on the chemical equilibration was performed~\cite{munshi} and it was
seen that the rapid transverse expansion of the plasma, leads to 
a rapid cooling and hence the chemical equilibration is considerably
impeded. In fact in the regions, where the radial velocities are
really large, the system may even move away from equilibration.

The treatment was extended to study the equilibration of 
strangeness, with a proper accounting of the mass of the strange quarks,
and once again it was found that the radial flow slows down the 
approach to strangeness equilibration~\cite{s_munshi}.

\subsection{Hydrodynamical Evolution of  QGP}

Space time evolution constitutes the most important aspects of
quark gluon plasma studies. Relativistic hydrodynamics although
classical in concept provides a computational tool to understand
at least the gross features of heavy ion collisions at ultra-relativistic
energies. Thermalization time scales mentioned above are required
as one of the inputs to solve the hydrodynamical equations.
Assuming a first order phase transition scenario,
(3+1) dimensional hydrodynamic equations with boost invariance
along the longitudinal direction have been solved to
estimate the space-time volumes of the QGP phase, mixed phase of QGP and
hadrons and the pure hadronic phase~\cite{q5}.
Transverse momentum
spectra of photons have been evaluated in the frame work
of hydrodynamical model~\cite{q4,q5}.
The effects of dissipation on the
space-time volume of the QGP has been explored and
found that the presence of dissipation reduces the rate of cooling,
resulting in longer total life time of the system~\cite{viscosity}.

A  general formulation of the relativistic hydrodynamics has also
been developed~\cite{q2}
which provides a bridge between the two extreme and largely idealized
scenarios of complete stopping earlier proposed by L. D. Landau and
longitudinal boost invariance
proposed by J. D. Bjorken and is thus better
applicable in the analysis of ultra-relativistic heavy ion collision
processes.

\subsection{Evolution of Fluctuation in Relativistic Heavy Ion Collisions}

The time evolution of the fluctuations in
the net baryon number for different initial conditions and space
time evolution scenarios have been considered.
It is  observed that the fluctuations
at the freeze-out depend crucially on the equation
of state (EOS) of the system and for realistic EOS
the initial fluctuation is substantially dissipated
at the freeze-out stage.
At SPS energies the fluctuations in net baryon number at the freeze-out
stage for quark gluon plasma and hadronic initial state
is close to the Poissonian noise for ideal as well as for EOS
obtained by including heavier hadronic degrees
of freedom. For EOS obtained
from the lattice QCD,  the fluctuation is larger than
Poissonian noise.  It is also observed that
at RHIC energies the fluctuations at the freeze-out point deviates
from the Poissonian noise for ideal as well as realistic equation of state,
indicating the presence of dynamical fluctuations~\cite{fluctuation}.

\subsection{Probes for the Equation of State}

Because of the extremely small life time and volume
it is not possible to measure directly any thermodynamic
properties of the QGP. Therefore, one has to look
for relations between the thermodynamic variables and
experimentally measurable quantities. Such possibilities are
to relate the entropy of the system with the measured multiplicity
and temperature with the average transverse momentum of the hadrons emanating
from the system. The variation of entropy (multiplicity) with
temperature (transverse momentum) will indicate the phase transition.

Therefore, the variation of average transverse mass of identified
hadrons with charge multiplicity have been studied for AGS, SPS and
RHIC energies. The observation of a plateau in the average transverse
mass for multiplicities corresponding to SPS energies is attributed to
the formation of a co-existence phase of quark gluon plasma and hadrons.
A subsequent rise for RHIC energies may indicate a de-confined phase in the
initial state.

Several possibilities which can affect the average transverse mass are
discussed~\cite{eosprobe}. Constraints on the initial temperature and
thermalization time have been put from the various experimental
data available at SPS energies.
It has been shown by solving the hydrodynamic equations that
the presence of mixed phase really slows down
the growth of  average transverse mass with increase in initial energy density.

\subsection{Spectral Change of Hadrons at High Temperature and Density}

\begin{figure}
\centerline{\psfig{figure=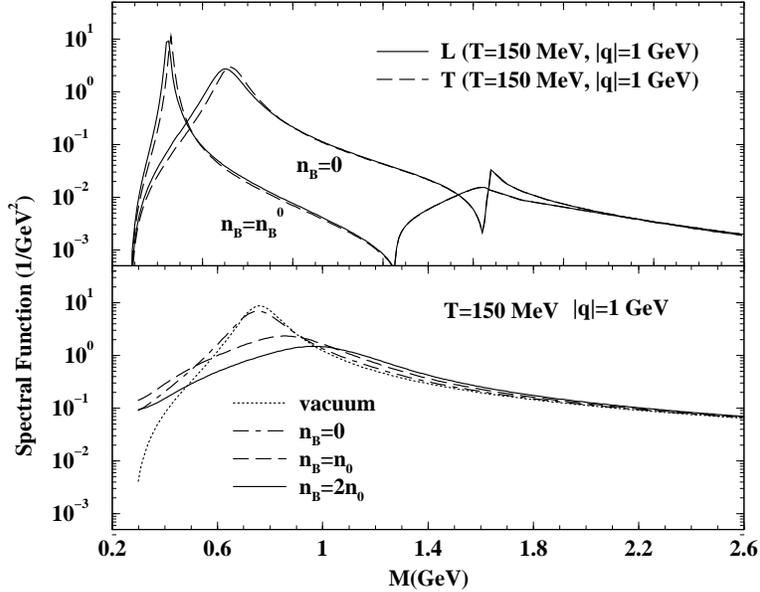,height=8cm,width=10cm}}
\caption{The spectral function of $\rho$ at a temperature ($T$) 150 MeV
and three momentum ($\mid q \mid$)  1 GeV for different baryonic densities.
}
\label{fig2_jane}
\end{figure}

The strongly interacting system formed after the nuclear collisions
provides a thermal bath where the spectral function of the hadrons
may be very different from its vacuum counterpart. Hence the
chiral symmetry  which is broken in the vacuum may be restored
in a hot and dense thermal bath. The changes in the hadronic
properties have been computed within the ambit of Quantum
Hadrodynamics (Fig.~\ref{fig2_jane})
and results are compared with gauged linear and non-linear
sigma models, hidden local symmetry approach,  QCD sum rule approach
\cite{photon-med,in-med,annals}.
Subtleties, such as the implications of the generalization of
Breit-Wigner formula for non-zero temperature and density,
question of collisional broadening, the role of Bose enhancement,
the possibility of kinematic opening (or closing) of decay channels due
to environmental effects have been studied in detail~\cite{photon-med}.

The change in hadronic properties
at finite temperature and density  is  important
not only to understand the restoration of chiral symmetry but
in such a situation the background for the signal of QGP will
also change. This has been explicitly demonstrated for the electromagnetic
probes of QGP~\cite{photon-med}.

\subsection{Medium Effects in Photon-Nucleus Interactions}

The heavy ion experiments and the corresponding theoretical studies
remain inconclusive on the nature of medium effects. This is
so mainly due to the fact that the medium effects
on hadrons are masked by complicated dynamics both in the initial
and final states. On the other hand, these difficulties are
largely overcome with the use of photons (as projectiles) which do
not have the problem of initial state interaction.

\begin{figure}
\centerline{\psfig{figure=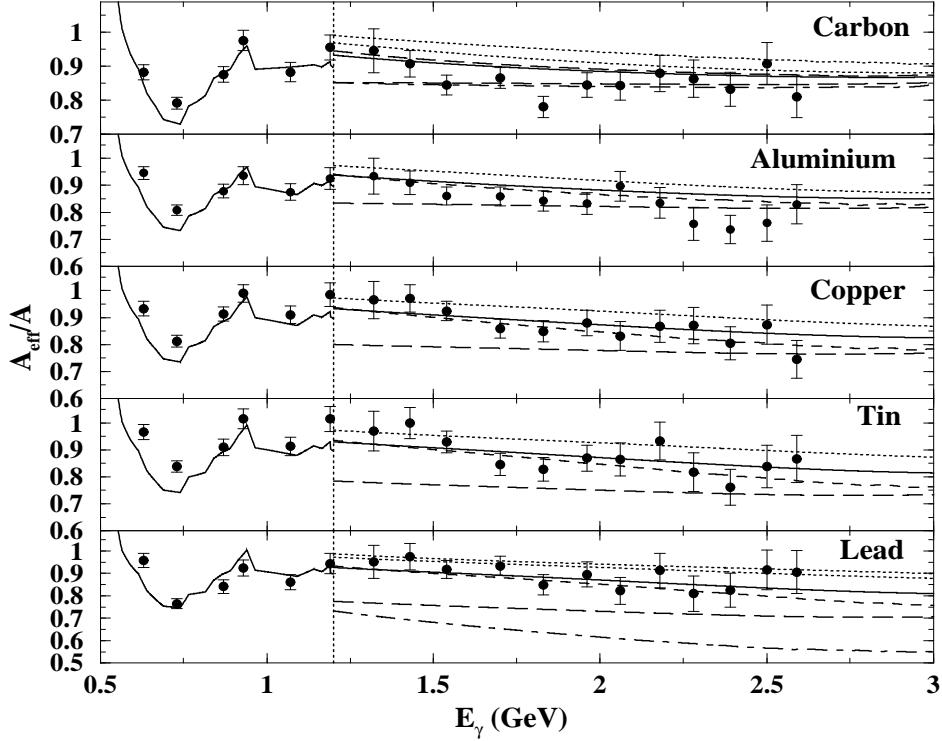,height=10cm,width=12.5cm}}
\caption{
$A_{\mathrm{eff}}/A$ for various nuclei as a function of photon
energy. For $E_\gamma\,<\,1.2$ GeV the results
for the baryonic resonance contribution are shown. For photon
energy $\geq 1.2$ GeV we show the results for both multiple
scattering approach and Glauber model. The dotted, long-dashed and
solid lines indicate calculations using Glauber model for vacuum, QHD and
USS respectively. The circles, dot-dashed (shown for C and Pb) and
short-dashed lines correspond to the same in the multiple scattering approach.
}
\label{fig3_jane}
\end{figure}

The  effects of in-medium hadronic properties
on shadowing in photon-nucleus interactions in Glauber model
as well as in the multiple scattering approach have been
studied and  it is found~\cite{photoabs1} that  the experimental data can
be reproduced  with the reduction of hadronic mass inside the nucleus
(Fig.\ref{fig3_jane}).

\begin{figure}
\centerline{\psfig{figure=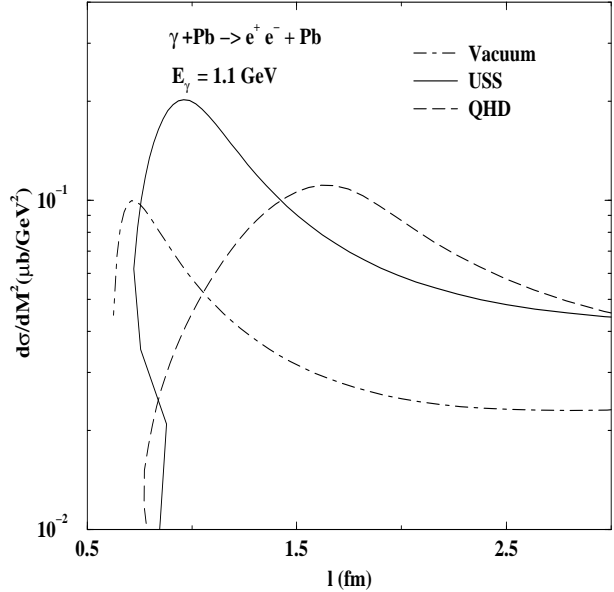,width=8cm,height=8cm}}
\caption{Invariant mass distribution of lepton pairs from $\gamma$-Pb
collisions at  $E_\gamma=1.1$ GeV. The result indicated by vacuum
corresponds to the mass of the vector meson in vacuum but in-medium
effects are included in the width.
The curves denoted by USS and QHD correspond to the
medium dependent masses given by universal scaling scenario
and quantum hadrodynamics respectively.
}
\label{fig4_jane}
\end{figure}

To observe the medium effects in photon-nucleus collisions
very clearly one may tune the incident
photon energy to 1.1 GeV~\cite{photoabs2} so that the $\rho$ meson is created inside
the nucleus at rest. In such a situation the $\rho$ meson is
forced to decay inside the nucleus and will signal the in-medium
effects  through dileptonic decay as seen in Fig~\ref{fig4_jane}.

\subsection{Electromagnetic Probes of Quark Gluon Plasma}

Among many signals of QGP proposed in the literature,
the electromagnetic probes {\it i.e.} photons and dileptons,
are known to be advantageous as these signals probe the
entire volume of the plasma, with little  interaction and
thus, are better marker of the space time history of the
evolving fireball created after the collisions~\cite{review}. In view of this
a detailed study was carried on the spectra of photons
and dileptons emitting from the strongly interacting system  formed
after the collisions of two nuclei at SPS and RHIC
energies.  An exhaustive set of reactions involving both
mesons~\cite{photon-med,piro}
and baryons~\cite{baryon} have been considered to evaluate
photon production rates from hadronic matter.

The formulation of the
production of photons and lepton pairs from QGP and hot hadronic
gas based on finite temperature field theory have been studied~\cite{annals}.
The changes in the
spectral functions of the hadrons appearing in the internal
loop of the photon self energy diagram have been considered in the
framework of Walecka model, gauged linear sigma model, non-linear sigma model,
hidden local symmetry approach
and universal scaling scenario.
The hadronic spectral
functions (in vacuum) for the isovector and isoscalar channel have been
constrained from the experimental data of $e^+e^-\rightarrow$ hadrons.
The effects of the continuum on the dilepton spectra are
included and seen to be substantial.
Relativistic  hydrodynamics has been used to describe the space time
evolution of the matter.
It is  observed that the in-medium effects on the hadronic
properties within the frame work of
the Gauged Linear and Non-Linear Sigma Model,
Hidden Local Symmetry approach are too small
to affect the electromagnetic spectra substantially.
However, the shift in the hadronic properties of
    different magnitude  within the frame work of the
Walecka model, universal scaling
scenarios  are prominently visible through the (low) invariant mass
distribution of dileptons and transverse momentum spectra
of photons.  Within the ambit of the present calculations
it is  observed that  the space-time integrated
photon spectra from the ``hot hadronic matter initial state''
outshine those originating from the ``QGP initial state'' for the
entire range of $p_T$, making it difficult to ``detect''
QGP via photon yield at SPS energies.
At RHIC energies, however, a scenario of a pure hot hadronic system
appears to be unrealistic because of the very high initial temperature.
It is observed that at RHIC energies the thermal photon (dilepton)
spectra originating from Quark Gluon Plasma over shines those from
hadronic matter for high transverse momentum (invariant mass) irrespective
of the models used for evaluating the finite temperature effects
on the hadronic properties.

As an important point of the present calculations it is
observed that the dilepton
spectra are affected both by the changes in the decay width as
well as in the mass of the vector mesons.
However, the photon spectra are affected
only by the change in the mass of the vector mesons but  are rather
insensitive to the change in its width~\cite{wa98}.
It is noted that Walecka model calculations
give different mass shift for $\rho$ and
$\omega$ mesons (because $\rho$ and $\omega$ couple to
nucleons with different strength). The disentanglement of the
$\rho$ and $\omega$ peaks in the dilepton spectrum resulting
from the ultra-relativistic heavy ion collisions
would be an excellent evidence of in-medium mass
shift of vector mesons and/or validity of
such model calculations for the situation under consideration.

\begin{figure}
\centerline{\psfig{figure=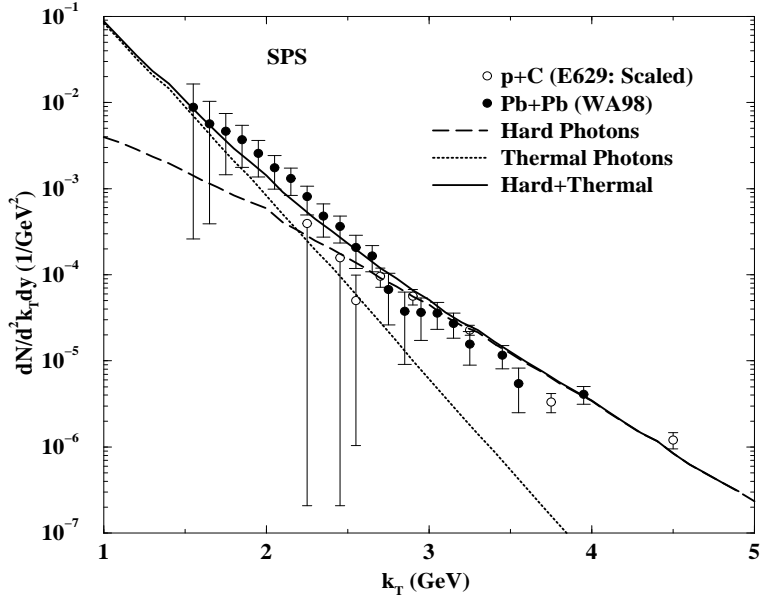,height=8cm,width=10cm}}
\caption{Single photon spectra from Pb + Pb collisions at 158A GeV.
}
\label{fig5_jane}
\end{figure}

The  photon spectra measured
by WA98 collaboration at CERN SPS energies in Pb + Pb collisions
has been studied by using the model described above.
The change in
the hadronic spectral function is taken into account both
in the production cross section of photon and equation
of state.
The WA98 photon data is well reproduced (Fig.~\ref{fig5_jane}) by both
hadronic as well as quark gluon plasma initial state
with temperature $\sim$ 200 MeV, indicating that the
data can not make firm conclusions about the formation of
quark gluon plasma~\cite{wa98}.  It has been noted that the WA98 data
can not be reproduced with large broadening of $\rho$.
Similar conclusion was drawn from the analysis of
photon spectra measured by WA80 collaboration in S + Au
collisions at SPS energies~\cite{wa80}. The effects of viscosity on
the space time evolution have been included in evaluating
the photon yield. It was found that the effect of dissipation
on the thermal photon is seen to be important in QGP as compared to
the hadronic phase~\cite{viscosity}.

\begin{figure}
\centerline{\psfig{figure=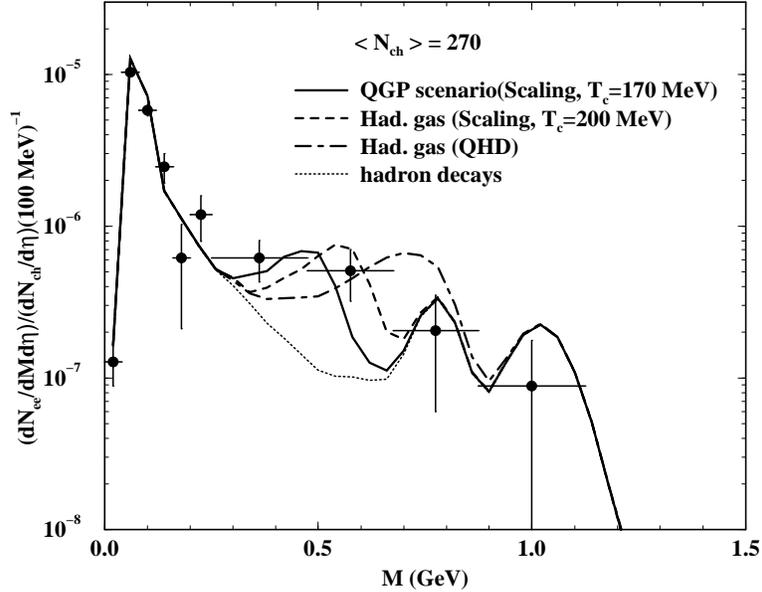,height=8cm,width=10cm}}
\caption{Invariant mass distribution of lepton pairs
 from Pb + Au collisions at 158A GeV.
}
\label{fig6_jane}
\end{figure}

The dilepton data obtained by CERES collaboration in Pb + Au
collisions at SPS energies have 
been analyzed in the same framework~\cite{ceres}.
Interestingly,
the data is well described (Fig.~\ref{fig6_jane}) by QGP and also by hadronic
initial state of initial temperature $\sim$ 200 MeV, when
the reduction in hadronic masses according to universal
scaling is incorporated.  The freeze-out conditions of
the fire ball have been constrained by the transverse
mass spectra of pions and protons~\cite{hydro2}.

\subsection{QCD Phase Transition in the Early Universe}
The QCD phase transition has important consequences in cosmology too.
The possible remnants that may have survived the primordial epoch till date
can provide valuable clues about the nature of the phase transition.
A first order QCD phase transition in the microsecond old universe
could lead to the formation of quark nuggets made of $u$, $d$ and
$s$ quarks at a density somewhat larger than normal nuclear matter
density. It has been shown that primordial quark nuggets with sufficiently
large baryon number could survive even today and could be a possible candidate
for
the baryonic component of cosmological dark matter~\cite{QN1}.
The abundance and size distribution of the quark nuggets
have been  evaluated~\cite{QN2} for different nucleation rates
proposed in the literature.
It is  found that there are a large number of stable quark nuggets which
could be a viable candidate for cosmological dark matter. The dependence
of the abundance and size of the quark nuggets on the value of the
critical temperature and the surface tension of
the quark matter have also been studied.

\subsection{$J/\psi$ Suppression in pA/AA Collisions}

In  relativistic  heavy  ion  collisions $J/\psi$ suppression has
been recognized as an important tool  to  identify  the  possible
phase transition to quark-gluon plasma (QGP). In a QGP due to
Debye screening, binding
of  $c\bar{c}$  pairs  into  a  $J/\psi$  meson is hindered and
$J/\psi$ production is  suppressed.

NA50 collaboration measured centrality dependence of
J/$\psi$ suppression in Pb+Pb collisions at 158 A GeV. Suppression
is more in central than in peripheral collisions and  is termed
anomalous as it goes beyond  the  conventional  suppression  in  a  nuclear
environment.

\begin{figure}
\centerline{\epsfig{figure=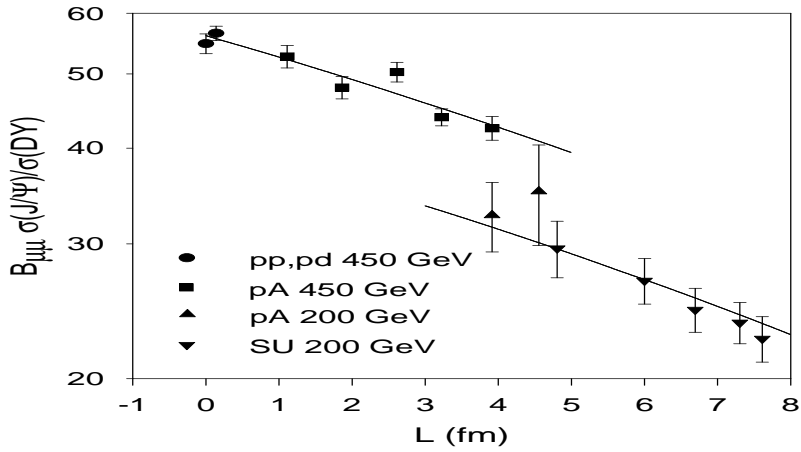,height=10cm,width=12.5cm}}
\vskip -3cm
\caption{
The experimental ratio of total $J/\psi$ cross section
 and Drell-Yan cross sections in proton-proton, proton-nucleus and
nucleus-nucleus collisions. The fit to the data obtained in the QCD
based nuclear absorption model is shown as solid lines.
 }
\label{akc1}
\end{figure}
\begin{figure}

\centerline{\epsfig{figure=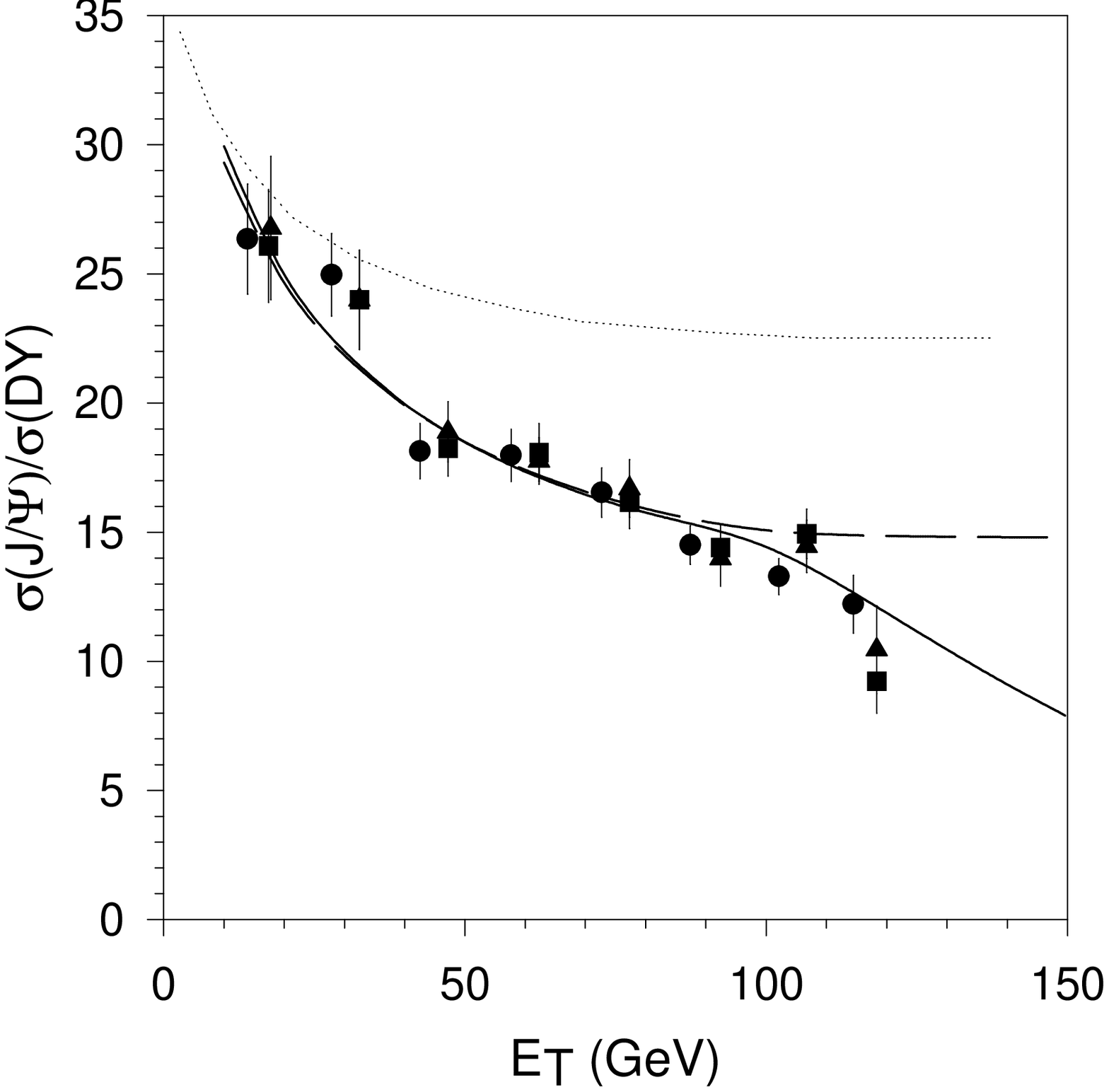,height=10cm,width=12.5cm}}
\vskip -3cm
\caption{
The transverse energy dependence of
$J/\psi$ over Drell-Yan ratio in 200 GeV/c S+U collisions. The dashed
and dotted lines are the fit obtained in the Glauber model of nuclear
absorption with $\sigma^{J/\psi N}_{abs}$=7.1mb and 4.4 mb respectively.
The solid line is the fit obtained in the QCD based nuclear absorption model.
 }
\label{akc2}
\end{figure}

In a series of papers \cite{jpsi} we have studied $J/\psi$ suppression in
nuclear medium.
We have proposed a QCD based nuclear absorption model to explain the anomalous $J/\psi$ suppression.
Production of $J/\psi$ meson is assumed to be a two step process,
(i) production of $c\bar c$ pairs with relative momentum square
$q^2$, and (ii) formation  of $J/\psi$ mesons from the $c\bar{c}$
pairs. Step (i) can be accurately calculated  in  QCD.
 The  second step, formation  of  $J/\psi$
mesons from initially compact $c\bar{c}$ pairs  is
non-perturbative. We  use a parametric form for the step (ii),
formation of $J/\psi$ from $c\bar{c}$ pairs, which is a function of the
relative square momenta $q^2$.
In  a nucleon-nucleus/nucleus-nucleus  collision,  the produced
$c\bar{c}$ pairs interact with nuclear medium before they  exit.
It is assumed that due to interaction with nuclear medium, $c\bar{c}$
pair gains relative square momenta. In traversing a length $L$ in the medium,
relative square momentum is changed, i.e., 
$q^2 \rightarrow q^2 +\varepsilon L$.
Square momentum gain per unit length was obtained from fitting
total $J/\psi$ production in pA/AA collisions. The model then reproduces the anomalous $J/\psi$ suppression in 158 AGEV Pb+Pb collisions. Representative
results are shown in Figs.~\ref{akc1} and \ref{akc2}.

How does the $J/\psi$ dissociate in the QGP? Does it dissociate due to 
the Debye screening of colour interaction and the consequent
dissolution of resonances? Or it dissociates due to scattering with
high energy gluons which may be present in the plasma? A comparative
study of the two mechanisms was done to locate the phase-space where
one of them dominates\cite{gjpsi}.

It was also found that the transverse momentum dependence of the
suppression factor for $J/\psi$ and $\Upsilon$ depends sensitively
on the speed of sound in the plasma, which controls the rate
of cooling of the plasma~\cite{jpsi_d}. This may help determine
the equation of state of the QGP.

\subsection{Disoriented Chiral Condensate}

Equilibrium  high  temperature QCD exhibit chiral symmetry if the
quarks are assumed to be  massless.  At  a  critical  temperature
$T_c$,  chiral  symmetry is spontaneously broken by the formation
of a scalar ($<\bar{q}q>$) condensate.    In  hadron-hadron  or  in  heavy   ion   collisions,   a
macroscopic region of space-time may be created, within which the
chiral  order  parameter is not oriented in the same direction in
the internal $O(4)=SU(2) \times SU(2)$  space,  as  in the  ordinary
vacuum.   Disoriented chiral condensation,
in hadronic or in heavy ion collisions can lead to  the
spectacular  events  that  some  portion  of the detector will be
dominated by charged pions or by neutral pions only. In contrast,
in a general event, all  the  three  pions  ($\pi^+$, $\pi^-$  and
$\pi^0$)  will  be  equally  well  produced.

If in a heavy ion collision, a certain
region undergoes chiral symmetry restoration, that region must be
in contact with some environment or background. Exact  nature  of
the  environment  is  difficult  to  determine  but presumably it
consists  of  mesons  and   hadrons   (pions,   nucleons   etc.).
Recognizing   the   uncertainty   in  the  exact  nature  of  the
environment, we choose to represent it by a white  noise  source,
i.e.  a  heat  bath. To analyze the effect of environment or heat
bath, on the possible disoriented chiral condensate,  we  propose
to  study  Langevin  equation for linear $\sigma$-model.

The linear sigma model Lagrangian can be written as,

\begin{equation}
{\mathcal{L}} =\frac{1}{2} (\partial_\mu \Phi)^2 -
\frac{\lambda}{4}
(\Phi^2 - v^2)^2 + H \sigma, \label{1}
\end{equation}
\noindent  where  chiral  degrees  of freedom are the O(4) fields
$\phi_a =(\sigma, \roarrow{\pi})$. In  Eq.\ref{1}  $H$  is
the explicit symmetry breaking term. This term is responsible for
finite  pion  masses. The parameters of the model, $\lambda$, $v$
and $H$ can be fixed  using  the  pion  decay  constant  $f_\pi$,
$\sigma$  and  pion  masses. With standard parameters, $f_\pi$=92
MeV, $m_\sigma$=600 MeV and $m_\pi$=140 MeV, one obtains,

$$\lambda=\frac{m_\sigma^2-m_\pi^2}{2f^2_\pi}       \sim      20,
v^2=f^2_\pi-\frac{m^2_\pi}{\lambda}=(87     MeV)^2,     H=f_\pi
m^2_\pi= (122 MeV)^3$$

We write the  Langevin  equation  for linear $\sigma$ model in ($\tau,x,y,Y$)
coordinates as,

\begin{equation}
[\frac{\partial^2}{\partial \tau^2} +(\frac{1}{\tau}+\eta)
\frac{\partial}{\partial \tau}
-\frac{\partial^2}{\partial x^2} -\frac{\partial^2}{\partial y^2} -
\frac{1}{\tau^2} \frac{\partial^2}{\partial Y^2}
+\lambda (\Phi^2 - f^2_\pi -T^2/2)] \Phi
 = \zeta (\tau ,x,y,Y)\label{1a}
\end{equation}

\begin{figure}[h]
\centerline{\psfig{figure=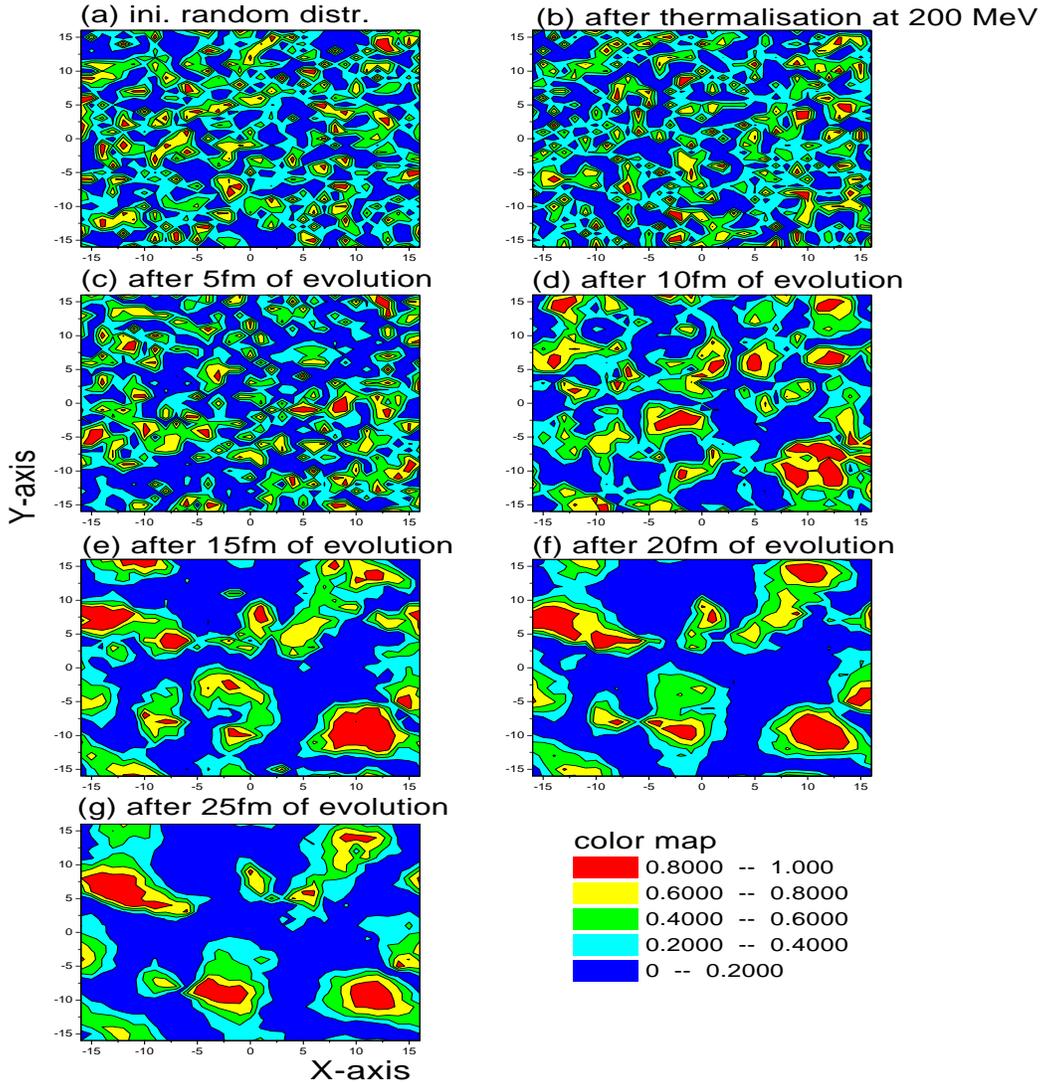,height=16cm,width=15cm}}
\caption{
Contour  plot  of the neutral to total pion ratio, at
rapidity Y=0. The explicit symmetry breaking term is omitted i.e.
pions are massless. The cooling law corresponds to  fast  cooling
law.  Different  panels  show  the  evolution  of  the  ratio at
different times. Domain like structure is evident at late times.
 }
\label{akc3}
\end{figure}

\begin{figure}
\centerline{\psfig{figure=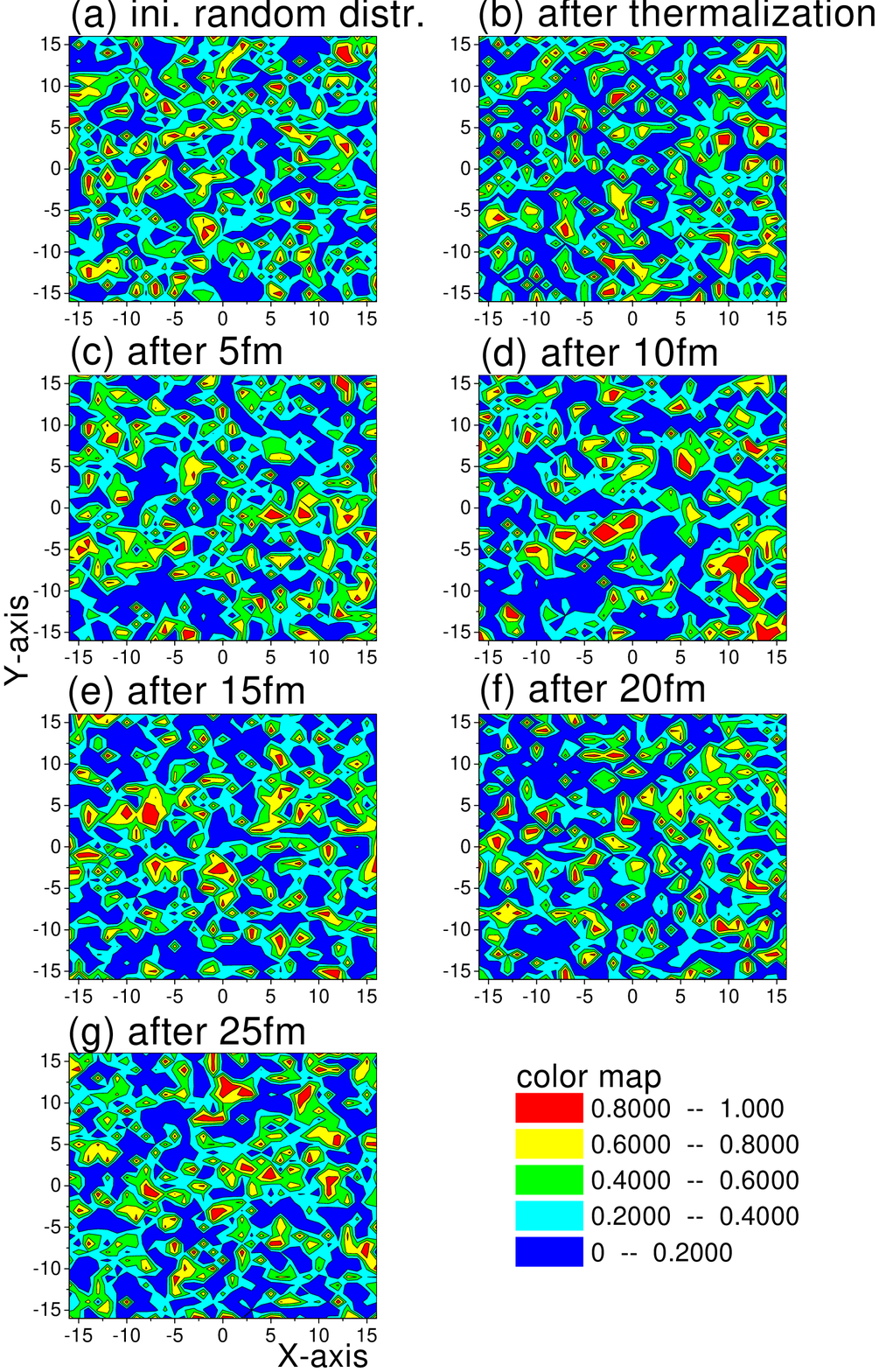,height=16cm,width=15cm}}
\caption{
Contour  plot  of the neutral to total pion ratio, at
rapidity Y=0. The explicit symmetry breaking term is omitted i.e.
pions are massless. The cooling law corresponds to  slow  cooling
law.  Different  panels  show  the  evolution  of  the  ratio at
different times. Domain like structure is evident  at  very  late
times
 }
\label{akc4}
\end{figure}

We have simulated the Langevin equation for linear sigma model on a
$32^3$ lattice with lattice spacing of 1 fm \cite{dcc}.
Initially
random fields are thermalized at a temperature of 200 MeV,
 above the critical temperature.
If thermalized fields,  are cooled down slowly, domains of
disoriented chiral condensates are not formed. On the other hand, if
thermalized fields are cooled
rapidly, and if the pions are {\em massless}, domains of disoriented
chiral  condensate  domains  are  formed,  quite  late   in   the
evolution. For massive pions, even in rapid cooling
disoriented chiral condensates are not formed.
 In Fig.\ref{akc3} and \ref{akc4}, temporal evolution of
the ratio $\pi^0/(\pi^++\pi^-+\pi^0)$, x-y plane, for massless and
massive pions are  shown.
Formation of domains of disoriented chiral condensate for massless pions
are evident.

\section{Parton Cascade Model}

One of the most ambitious treatments for the relativistic heavy 
ion collision of nuclei is provided by the parton cascade model developed
at VECC in collaboration with the Duke group. In these calculations
we treat the nuclei as consisting of valence and sea quarks and
gluons, which  scatter, radiate, and fuse according to the
cross-sections determined from perturbative QCD. The attendant
singularities in the cross-sections are avoided by introducing a
lower cut-off in  the momentum transfer and in the virtuality,
below which the partons do not radiate.

Using this treatment, several very important results have 
been obtained, which include the energy density, the 
Debye screening mass\cite{pcm1}, the strangeness production\cite{pcm2},
 the net baryon
production~\cite{pcm3}, and the single photon production
\cite{pcm4}. 

In the course of these studies it was also noted that a jet of
quarks passing through quark gluon plasma should produce high
energy photons. At LHC energies this may be the large source of 
photons having a large transverse momenta\cite{fms} and these
should also be reflected in dilepton spectra~\cite{fms_dil}.

Dilepton tagged jets\cite{jets} is another development which 
promises to become very useful to precisely determine the
rate of energy loss of jets traversing the QGP.

\section*{Apologies and Acknowledgments}

The work discussed here is by no means exhaustive. A lot of high quality
work, spread over more than
100 research publications
 has gone unrepresented and is
not even referred, while several seminal papers have been
discussed very briefly, partly because they are by now very well
known. Several firsts, like estimates of energy loss of heavy
quarks while passing through a  quark gluon plasma, the diffusion of heavy 
quark in the QGP, theta vacuua, a large body of work on 
multiplicity distribution of particles produced in nuclear collisions,
a large number of papers on multi-fragmentation and liquid gas
phase transition,
a huge body of work on nuclear optical model potential and inelastic
scattering, application of the parton cascade model to SPS energies,
several bench mark calculations of prompt photon and Drell Yan 
calculations, some empirically determined scaling laws for particle
production, a very cute idea to identify
collective flow in nuclear collisions, several outstanding 
papers on transport theories and signal processing, etc.,
 etc., etc.,  have not
found place in this write-up as we got scared of the bulk!
Several experimentalists of our group have used the clarity of their
thought process to write valuable papers on fluctuations, fractals,
wave-lets, electronic stricture of atoms, nuclear structure, etc.
which could not be discussed.

It has not been possible to  even mention 
the hundreds of talks given by our members at
various conferences and courses taught
by them which have established schools of research and have led to
advancement of knowledge. The nuclear theory group at VECC has also
taken a leading role in organizing several schools and conferences.

We are grateful to our students and post doctoral
fellows; 
 Ananya Das,
Champak Baran Das,
 Piyushpani Bhattacharjee, 
Abhijit Bhattacharya,
Rupayan Bhattacharya,
 Somenath Chakraborty,
Sanjay Ghosh,
V. S. Uma Maheswari, 
 Sheela Mukhopadhyay,
Munshi Golam Mustafa, 
 Dipali Pal,
Binoy Krishna Patra,
Pradip K. Roy,
Tapas Sil, 
 and many others who
contributed to our efforts and now occupy responsible positions
across the world.

We are grateful to our collaborators,
 Steffen A. Bass,
 Debbrata Biswas,
Jean C. Cleymans,
Subal Das Gupta,
Bhaskar Dutta, 
 Klaus Geiger,
Rainer J. Fries,
Charles Gale,
Sean Gavin,
D. H. E. Gross,
Mikolos Gyulassy,
 T. Hatsuda,
Ulrich Heinz,
 Peter E. Hodgson,
 Joe Kapusta, 
Larry McLerran, 
 Berndt M\"{u}ller,
Shankar N. Mukherjee, 
Lakshmi N. Pandey, 
 Shashi C. Phatak, 
Sibaji Raha, 
 Heinigerd Rebel, 
Krzysztof Redlich,
Binayak Dutta Roy,
 Santosh K. Samaddar,
Chhanda Samanta,
Helmut Satz,
Radhe Shyam,
S. Shlomo,
D. Sperber,
Xin -Nian Wang,
 and many others who have collaborated 
with us over years and brought about a richness of intellectual
vibrancy to our group. That of-course is the greatest advantage
of pursuing physics as a career. 

And before we close, no theoretical research with predictive
power is possible unless good quality computational facilities are
available. We are privileged to have an access to continuously
improving computational facilities available at VECC.

We take this opportunity to put on record our most
sincere thanks to Dr. Bikash Sinha, who has been associated with
the theory group of the VECC from the very beginning and who has
shaped it to its present state of excellence. His leadership,
vision, and enthusiasm has been with us and will continue to guide
us.

Dr. Sinha always led us from the front and we consider ourselves 
very privileged to have `played' under his `captaincy'. He of-course
has played a true `captain's knock', all along.

\end{document}